\documentclass[fleqn,usenatbib]{mnras}

\usepackage{newtxtext,newtxmath}

\usepackage[T1]{fontenc}
\usepackage{ae,aecompl}

\usepackage{graphicx}	
\usepackage{amsmath}	
\usepackage{color}      
\usepackage{float}
\usepackage{hyperref}
\usepackage[normalem]{ulem}
\usepackage{arydshln}

\newcommand{\kms}{\text{km} \, \text{s}^{-1}}

\newcommand{\Ropt}{R_{\rm opt}}

\newcommand{\Vsm}{V_{\rm sm}}
\newcommand{\MHI}{M_\text{\HI}}
\newcommand{\Sint}{S_{\rm int}}
\newcommand{\rms}{\sigma_{\rm RMS}}
\newcommand{\rmsin}{\sigma_{\rm input}}
\newcommand{\wtw}{\omega_{20}}

\newcommand{\SN}{S/N}

\newcommand{\Afr}{A_{\rm fr}}
\newcommand{\Ncell}{N_{\rm cell}}
\newcommand{\NcellHI}{N_\text{cell,\HI}}
\newcommand{\HI}{H\,{\sc i}}
\newcommand{\HH}{H$_{2}$}

\newcommand{\Msun}{{\rm M}_{\sun}}
\newcommand{\lgMstar}{\log(M_{\star}/\Msun)}
\newcommand{\lgMstarMsun}{\log_{10} (M_{\star}/\Msun)}

\newcommand{\lgMh}{\log({\rm M}_{h}/\Msun)}
\newcommand{\lgMHI}{\log(M_{\text{\HI}}/\Msun)}
\newcommand{\Rvir}{R_\text{vir}}
\newcommand{\cp}{\citep}
\newcommand{\ct}{\citet}

\title[Global \HI\ asymmetries in IllustrisTNG]{Global \HI\ asymmetries in IllustrisTNG: a diversity of physical processes disturb the cold gas in galaxies}

\author[A. B. Watts et al.]{Adam B. Watts,$^{1,2}$\thanks{E-mail: adam.watts@research.uwa.edu.au}
Chris Power,$^{1,2}$ 
Barbara Catinella,$^{1,2}$ 
Luca Cortese,$^{1,2}$ and 
\newauthor{Adam R.~H. Stevens$^{1,2}$}
\\
$^{1}$International Centre for Radio Astronomy Research, The University of Western Australia, Crawley, WA, Australia\\
$^{2}$ARC Centre of Excellence for All-Sky Astrophysics in 3 Dimensions (ASTRO3D)\\
}

\date{Accepted 2020 October 09. Received 2020 October 08; in original form 2020 July 24}

\pubyear{2020}

\begin{document}
\label{firstpage}
\pagerange{\pageref{firstpage}--\pageref{lastpage}}
\maketitle

\begin{abstract}
Observations of the cold neutral atomic hydrogen (\HI) in and around disc galaxies have revealed that spatial and kinematic asymmetries are commonplace, and are reflected in the global \HI\ spectra. We use the TNG100 box from the IllustrisTNG suite of cosmological simulations to study the conditions under which these asymmetries may arise in current theoretical galaxy formation models. We find that more than 50\% of the sample has at least a 10\% difference in integrated flux between the high- and low-velocity half of the spectrum, thus the typical TNG100 galaxy has an \HI\ profile that is not fully symmetric. We find that satellite galaxies are a more asymmetric population than centrals, consistent with observational results. Using halo mass as a proxy for environment, this trend appears to be driven by the satellite population within the virial radius of haloes more massive than $10^{13}\ \Msun$, typical of medium/large groups. We show that, while the excess of \HI\ asymmetry in group satellites is likely driven by ram pressure, the bulk of the asymmetric \HI\ profiles observed in TNG100 {are} driven by physical processes able to affect both the central and satellite populations. Our results highlight how asymmetries are not driven solely by environment, and multiple physical processes can produce the same asymmetric shape in global \HI\ spectra.
\end{abstract}



\section{Introduction}
In order to understand the formation and evolution of galaxies we need to understand the processes that influence their cold ($T<10^4$\,K) gas reservoirs. These reservoirs fuel star formation and active galactic nuclei (AGN), and primarily exist in the form of neutral atomic hydrogen (\HI), which can be observed though its 21-cm emission line. When spatially integrated over a galaxy, the emission line spectrum (from here, the `global' \HI\ spectrum) corresponds to the \HI-mass-weighted line-of-sight velocity distribution of the \HI\ gas reservoir, thus encoding combined information about both the distribution and kinematics of the gas. As the \HI\ reservoirs of galaxies are typically detectable out to 2--3 times the radius of the optically bright stellar component \cp[e.g.][]{bigiel12}, global \HI\ spectra are sensitive to the dynamics in regions where baryons are less gravitationally bound.

Observations of global \HI\ spectra have revealed that they are typically not symmetric profiles \cp[e.g.][]{richter94,haynes98,watts20}, indicating that disturbances in the distribution and/or kinematics of the gas reservoir of a galaxy are common \cp[e.g.][]{sancisi76,swaters99,reynolds20,deg20}. The relative contribution to global \HI\ asymmetry from the different processes which govern galaxy evolution remains unknown, but we might expect there to be an environmental dependence because of the potential impact of processes such as gravitational tides and ram-pressure stripping. Galaxies in close pairs \cp{bok19} and in clusters \cp{scott18} appear to have more asymmetric global \HI\ spectra, as populations, than isolated galaxies. \ct{watts20} presented clear evidence that satellite galaxies are a more global \HI\ asymmetric population than centrals, and that global \HI\ asymmetric galaxies are gas-poor compared to symmetric ones at fixed stellar mass. This suggests that environmental processes, primarily those responsible for gas removal, are a dominant driver of asymmetries, at least in the stellar mass-selected and gas fraction-limited survey xGASS \cp[the extended GALEX Arecibo SDSS Survey;][]{catinella18}.

While it is clear that environmental processes can drive asymmetry, it remains unclear whether there is a dominant process and whether global \HI\ asymmetry measurements can be used to trace evidence of this. Current observational datasets do not allow us to fully address these questions. Blind \HI\ surveys such as ALFALFA \cp{haynes18} detect upward of 30 000 objects, but are restricted to the most gas-rich objects in their volume. Further, there are only $\sim$500 spatially resolved observations of the \HI\ in galaxies \cp{wang16}, which are similarly biased toward gas-rich objects. The environmental processes that remove and disturb the gas in galaxies are primarily gravitational and hydrodynamical, making them inherent and measurable in cosmological simulations of galaxy evolution \cp[e.g.][]{marasco16,yun19}. They resolve thousands of galaxies for which we can measure spatially resolved and integrated properties, along with accurate environment information provided by their 6D coordinates, making them a useful laboratory for testing the interpretation of our observations against our current theories of galaxy evolution. 

In this work we use one of the IllustrisTNG \cp{nelson19} cosmological simulations to explore how a global \HI\ asymmetry parameter is influenced by environment using metrics typically adopted in simulations: in terms of satellites and centrals and as a function of halo mass. The outputs of the simulation are visually inspected to investigate the connection between the asymmetry parameter and the resolved gas reservoir, to clarify how useful it might be as a tracer of environmental processes, and this is placed in context of future \HI\ surveys. 

This paper is organised as follows: in \S\ref{sec:sample} we describe our sample of IllustrisTNG galaxies, in \S\ref{sec:methods} we describe our asymmetry measurements and mock global \HI\ spectrum creation, and in \S\ref{subsec:noise} we investigate the effect of resolution on asymmetry measurement. In \S\ref{sec:results} we present the rate of global \HI\ asymmetry in the simulation and its variation with environment, in \S\ref{sec:discussion} we discuss the relationship between gas morphology and global \HI\ asymmetry, and in \S\ref{sec:concl} we conclude.

\section{Sample} \label{sec:sample}
We use data from one volume of the IllustrisTNG \cp[from here TNG,][]{pillepich18b,springel18,nelson18,naiman18,marinacci18,nelson19} cosmological magneto-hydrodynamical simulations,  which were run using the simulation code {\sc arepo} \cp{springel10} to solve the equations of magneto-hydrodynamics for gas on a moving Voronoi mesh, and Poisson's equation for gravity using the tree-particle-mesh method \cp{xu95}.
The simulation initial conditions were set using the \ct{planck} results: $\Omega_m = 0.3089$, $\Omega_\Lambda = 0.6911$, $\Omega_b = 0.0486$, $h = 0.6774$, $\sigma_8 = 0.8159$, and $n_s = 0.9667$ assuming a flat universe governed by a $\Lambda-$cold dark matter ($\Lambda$CDM) cosmology. 
In order to describe astrophysical processes occurring below the resolution limit of the simulations, TNG incorporates sub-grid models for stellar and AGN feedback, gas cooling, star formation, and massive black hole growth as described in \ct{weinberger17} and \ct{pillepich18a}. 
We use the TNG100 simulation, which has a periodic box of side length 75/$h$ cMpc and was initialised with $1820^3$ dark matter particles and gas cells.
Dark matter particles have a mass of  $7.5\times10^6\, \Msun$ while the average gas cell, and the  stellar particles formed from them,  have  $\sim1.4\times10^6\, \Msun$. 

The structure finding algorithm {\sc subfind} \cp{springel01,dolag09} was used to identify smooth, gravitationally bound collections of dark matter particles, referred to as haloes, with a friends-of-friends linking length equal to 0.2 times the mean inter-particle distance. 
{\sc subfind} also finds gravitationally bound overdensities inside haloes, labelled subhaloes. 
The most massive subhalo within each halo is designated the `central' and associated with the smooth, diffuse component of the halo while all others are designated `satellites'. 
We select 16377 subhaloes (hereafter referred to as `galaxies') with stellar-masses $\lgMstar > 9$ and  \HI-masses of  $\lgMHI>8$  \cp[i.e., a sub-sample of that used in][]{stevens19} {from the redshift zero snapshot of TNG100} to replicate the region of the \HI\ gas-fraction scaling relations covered by xGASS. 
In \S\ref{subsec:spec} we remove 74 galaxies with unphysical \HI\ profiles and a further 5604 in \S\ref{subsec:restest} with our quality cut, making 10699 the final number of galaxies studied in this work.

\section{Methods} \label{sec:methods}

\subsection{Simulated \HI-line profiles} \label{subsec:spec}

We analyse \HI\ spectral lines derived from the `inherent' TNG100 properties presented in \ct{stevens19}, and briefly describe their relevant methods here in addition to the methods specific to this work. 
{Within each subhalo, we selected a subset of the gravitationally bound gas cells using a spherical aperture with radius defined using the method of \ct{stevens14}: where the gradient of the cumulative baryonic (stars + neutral gas) mass profile becomes constant (i.e. an isothermal profile).}

\begin{figure*}
    \centering
    \includegraphics[width= 0.9\textwidth]{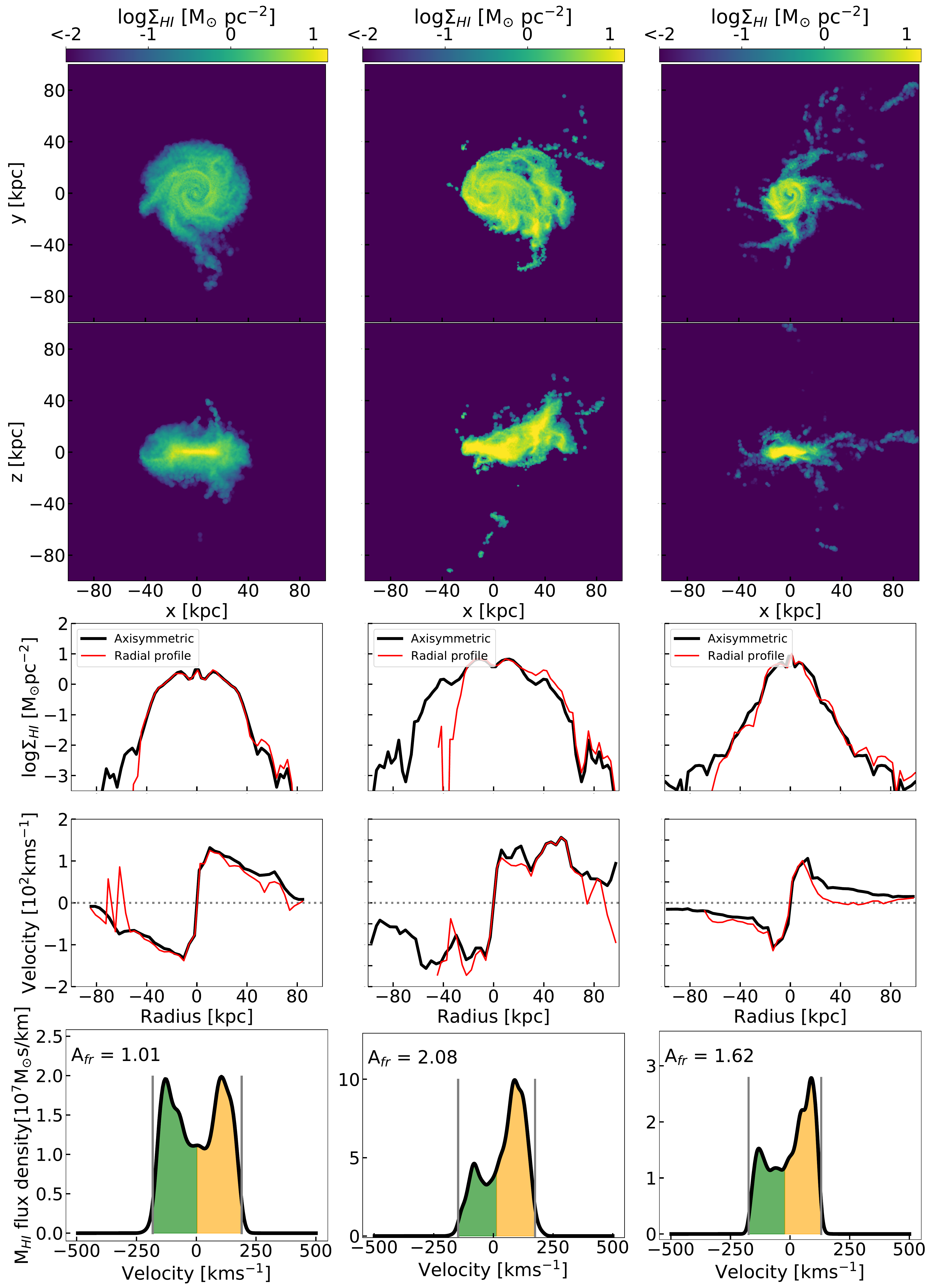}
    \caption{Example galaxies from TNG100. The first two rows are the resolved surface density of \HI\ in the face-on and edge-on projection, where the \HI\ mass of each cell has been distributed over a (projected) sphere with volume equivalent to the cell's. The density scale is given by the color bar above the top panels. The third and fourth rows are the radial face-on \HI\ surface density distribution and LOS velocity profile for the axisymmetric case (thick, black line) and the approaching (negative radius) and receding (positive radius) sides separately (red thin line). The `inherent' global \HI\ spectrum is shown in the last row with measurement limits (thin grey lines) and shaded regions (green and orange) used to calculate $\Afr$, which is shown in the top left of the panel.}
    \label{fig:spec_examples}
\end{figure*}

{This separates the main baryonic component of the halo (the galaxy) from the diffuse component (e.g. the circumgalactic medium).}
{The apertures are generally larger for more massive galaxies, and typical aperture sizes are sub 30\,kpc for galaxies with $\lgMstar\lesssim10.5$, but for the largest galaxies can be of order 100\,kpc.}
The neutral hydrogen fraction of non-star-forming gas cells is calculated in TNG100 by simulating the ionising photon background rate of \ct{faucher09} and accounting for self shielding \cp{rahmati13} and gas cell cooling rates. 
In star-forming cells where the hydrogen number density is above ${\rm n}_{\rm H} = 0.106\, {\rm cm}^{-3}$ the adopted \ct{springel03} star formation model partitions the hydrogen abundance into a two-phase `cold cloud' and `ambient hot' medium, and we assume that the entire cold component is neutral \cp[see][]{diemer18,stevens19}. 
The partition between \HI\ and \HH\ is then modelled using eq. 8 of \ct{gnedin14} as described in A.2 of \ct{stevens19}, where the UV field is modelled following the method described by \ct{diemer18}. 
Briefly, \ct{gnedin14} give formulae for calculating the neutral hydrogen surface density where the \HH/\HI\ ratio is unity based on the local UV radiation field and dust mass. 
The \HH/\HI\ fraction of each cell is then given by the ratio of the cell neutral hydrogen column density to the density calculated for unity. 
Adopting the inherent $z-$direction in the simulation as the line-of-sight (LOS), the spectra were created by summing the \HI\ mass of the cells in 2 $\kms$ bins of LOS velocity relative to the galaxy's centre of mass velocity, and smoothing by a Gaussian of width given by the \HI -mass-weighted mean LOS thermal velocity dispersion of the gas cells.

We note that using the `inherent' calculation, \ct{stevens19} found that galaxies in the stellar mass range $10.5<\lgMstar< 11$ {can} have a $\sim$1 dex lower \HI\ fraction than seen in observations, and that this primarily affects satellite galaxies, along with a small number of centrals (see their fig. 4). This has been attributed to AGN feedback as this stellar mass range is where the dominant feedback mechanism is known to change from stellar to AGN, and  \ct{pillepich18b} linked reduced gas fractions of TNG100 haloes with $\lgMh>12$ explicitly to AGN feedback. The \HI-mass cut we adopt in \S\ref{subsec:samp} will remove the majority of these AGN-affected galaxies due to their inherently low \HI\ fractions, and we have checked that our results are robust to harsher cuts. 

\begin{figure*}
    \centering
    \includegraphics[width= \textwidth]{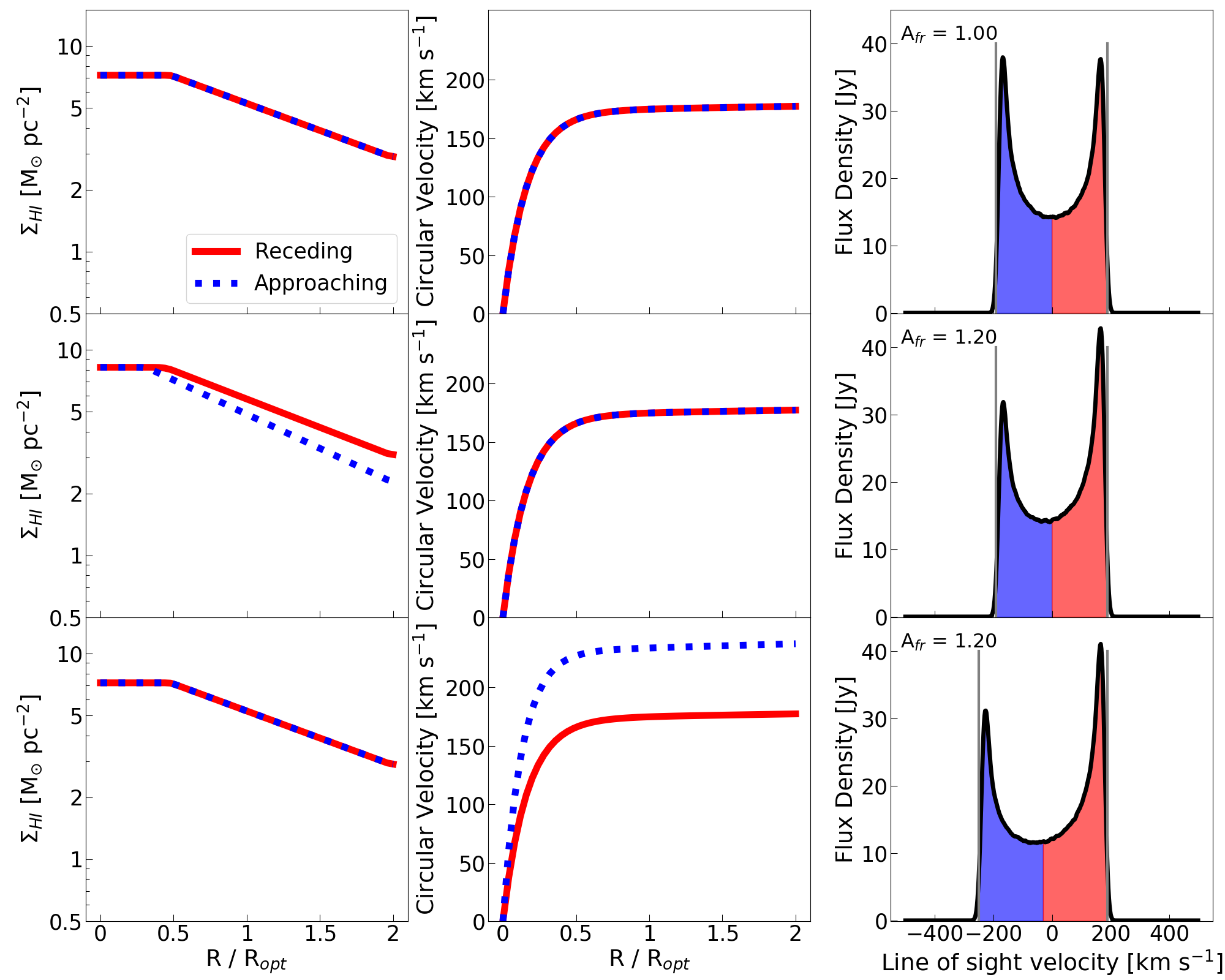}
    \caption{Demonstration of global \HI\ asymmetry measurement. Panels show the radial \HI\ surface density distribution (left), circular velocity profile (middle) and  global \HI\ spectrum (right) input into the model. The top row is a symmetric model, the middle row has an asymmetric \HI\ distribution  and the bottom row has an asymmetric rotation curve. The input \HI\ distribution and rotation curve on the approaching (negative line-of-sight velocity) and receding (positive line-of-sight velocity) sides are shown as solid-red and dotted-blue lines respectively. The model radii are normalised by $\Ropt$, defined as the radius enclosing 83\% of the integrated stellar light, as it is the normalisation adopted by the \ct{catinella06} RC templates. Measurement limits are shown on the \HI\ spectrum with vertical grey lines and the integrated flux in each half used to measure $\Afr$ is shaded blue and red.}
    \label{fig:Afr_demo}
\end{figure*}

We identify the edges of the emission-line spectrum by iterating along the velocity channels from the peak spectral flux density channel until it equals 20\% of the peak value (from here, the $\wtw$ measurement limits). 
The spectrum is split in half using these limits and the highest peak in each respective half is identified, then the $\wtw$ measurement limits are recalculated using the heights of the new peaks.
If there is no peak found in one half, the $\wtw$ measurement limits do not agree when approaching from their left and right sides, or there is $>40\%$ of the spectrum's total flux located outside the $\wtw$ limits, then we flag the spectrum for manual inspection.
During manual inspection we kept all single-peaked spectra using the peak value for both $\wtw$ limits, and removed {74} galaxies with unphysical spectral lines, leaving us with 15294 galaxies. 
{These unphysical spectral lines are primarily from $\lgMstar>10.5$ centrals in $\lgMh>12.5$ haloes with no clear \HI\ disc but significant, clumpy \HI\ throughout the halo.}
{This is likely due to AGN feedback, as discussed above, where the gas removed from satellites is now associated to the central and results in no distinct spectral line.}


In Fig. \ref{fig:spec_examples} we show the surface density of \HI\ in face-on (first row) and edge-on (second row) projections, radial face-on surface density distributions (third row) and edge-on rotation curves (RCs, fourth row) computed for the azimuthally averaged (axisymmetric) case and for the approaching ($x<0$) and receding ($x>0$) halves of the galaxy, and the global \HI\ spectrum (last row) for three examples of TNG100 galaxies. The leftmost galaxy is symmetric, and the mean deviations from the axisymmetric case for the receding and approaching halves, respectively, are 0.004 dex and -0.04 dex for the \HI\ distribution and  -7 $\kms$ and 6 $\kms$ for the RC. This results in a visually symmetric global \HI\ spectrum. The middle galaxy shows a diffuse, warped extension of gas on one side, characteristic of a  galaxy undergoing ram-pressure stripping by the intergalactic medium \cp[e.g. NGC 4522,][]{vollmer04}. The \HI\ distribution is clearly truncated on the approaching side, and the mean deviations for the receding and approaching halves, respectively, are 0.2 dex and -1.1 dex for the \HI\ distribution, and -10 $\kms$ and 10 $\kms$ for the RC. This indicates the disturbance is primarily in the \HI\ distribution and results in a global \HI\ spectrum with a suppressed peak. In the rightmost galaxy the \HI\ distribution appears regular while the receding RC drops quickly, and the mean deviations for the receding and approaching halves, respectively, are 0.02 dex and -0.07 dex for the \HI\ distribution, and -23 $\kms$ and 6 $\kms$ for the RC. This indicates that the disturbance is present primarily in the kinematics, giving rise to a global \HI\ spectrum with a higher peak in the half with slower rotation as the \HI\ spans a smaller range of LOS velocity.

\subsection{Asymmetry measurement} \label{subsec:asym}

The \HI-mass flux density ($ S_{\rm v}$) in each channel of a spectrum depends on the total \HI\ mass in a given LOS velocity channel, making it sensitive to both the \HI\ mass distribution and the kinematics of the gas, and also inclination. To quantify asymmetry we adopt the areal flux ratio parameter $\Afr$ \cp{haynes98} as it is the most frequently adopted parameter in the literature \cp[e.g.][]{espada11, scott18, bok19, reynolds20, watts20}. It is defined as
\begin{equation} \label{eq:Afr}
    \Afr = 
        \begin{cases}
            A & A \geq 1\\
            1/A & A < 1
        \end{cases},
\end{equation}
where $A$ is the ratio of the integrated flux in each half of the spectrum bounded by the limits $V_{\rm max}$ and $V_{\rm min}$ and divided by the middle velocity $V_M = 0.5(V_{\rm min} + V_{\rm max})$:
\begin{equation} \label{eq:A}
    A = \frac{
        \int_{V_{M}}^{V_{\rm max}} S_{\rm v} {\rm dv}
        }{
        \int_{V_{\rm min}}^{V_{M}} S_{\rm v} {\rm dv}
        }.
\end{equation}
We define the limits to be the $\wtw$ measurement limits, and also define the velocity width separating these limits as the 20\% velocity width ($\wtw$). 

It is worth mentioning that in a perfectly symmetric spectrum $V_M$ will correspond to the centre-of-momentum (or systemic) velocity of the system.
This quantity can be calculated explicitly for each galaxy in the simulation, but it can only be estimated observationally and can be sensitive to asymmetries in the spectrum. 
While it would be possible to measure $\Afr$ using the known centre-of-momentum velocity of each simulated galaxy, we adopt our $\wtw$  measurement limits and our eq. \ref{eq:Afr} definition to be consistent with how it would be measured in observational datasets.
{Another source of uncertainty associated with $\Afr$ is projection effects \cp[e.g.][]{deg20}.}
{If the asymmetric feature is aligned close to the projected minor axis of a galaxy, the measured $\Afr$ can be smaller than the intrinsic value, or even appear to be perfectly symmetric ($\Afr=1$).}
{This projection dependence makes $\Afr$ effectively a lower limit on the true asymmetry.}

In Fig. \ref{fig:Afr_demo} we demonstrate how asymmetries in the \HI\ surface mass density (middle row) and RC (bottom row) of a galaxy affects the $\Afr$ parameter for a double-horn spectrum created using the toy model described in \ct{watts20}. In the top row we show a symmetric galaxy with $\Afr=1$: the peaks and edges of the profile are the same on both sides, $V_M$ (located at the overlap of the red \& blue shading) is at $0\ \kms$ at the centre of the profile, and there is equal \HI\ integrated flux on each side of the galaxy. In the middle row the \HI\ distribution of the galaxy is more extended in one half: the \HI\ mass surface density is lower at a given radius in the approaching half (negative line of sight velocity) than the receding half. This leads to a suppression of the left side peak and a lower integrated flux under the blue shaded area, and an asymmetry value of $\Afr=1.2$. While there is a small change in $V_M$ as the lower left peak causes a shift in the left side velocity limit, the straight edges of the profile mean that the left velocity limit is almost unchanged. In this case, $\Afr$ traces the unequal amounts of \HI\ mass on either side of the galaxy, analogous to the middle galaxy in Fig. \ref{fig:spec_examples}. In the bottom panel the RC on the approaching side flattens 50 $\kms$ higher than the receding side, distributing the \HI\ over a larger velocity range. This shifts the left side velocity limit, and therefore $V_M$, to more negative velocities causing the blue shaded area to contain $<50$\% of the integrated flux and an asymmetry measurement of $\Afr=1.2$. In this case $\Afr$ traces the unequal RC, and is analogous to the rightmost galaxy in Fig. \ref{fig:spec_examples}.

\section{Resolution tests} \label{subsec:noise}
\subsection{Sampling uncertainty} \label{subsec:restest}
\begin{figure}
    \centering
    \includegraphics[width= 0.5\textwidth]{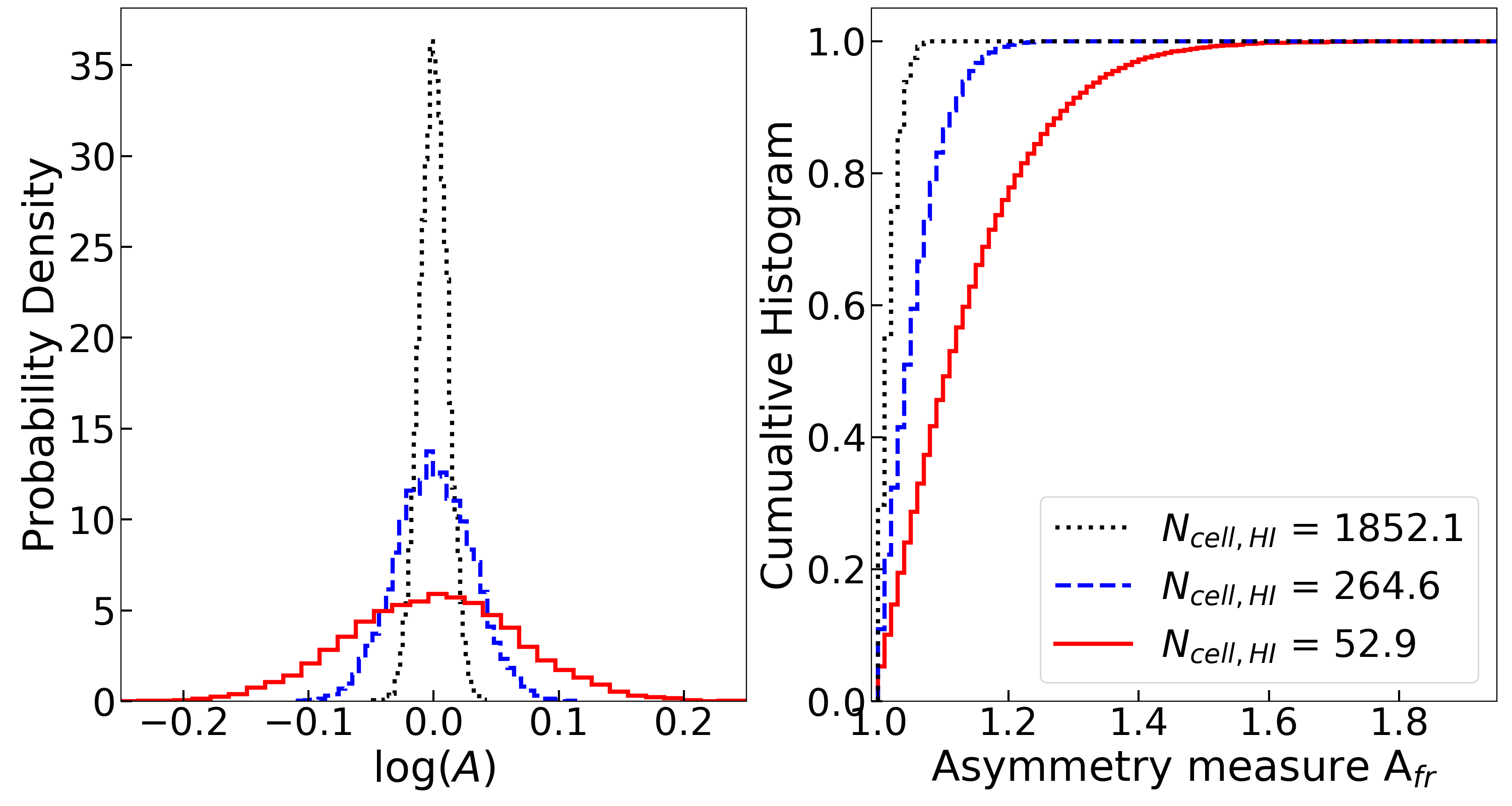}
    \caption{The effect of sub-sampling of gas cells on the measurement of global \HI\ asymmetry. Normalised histograms of recovered $\log(A)$ (left) and cumulative $\Afr$ distributions (right) from the $10^4$ sampling iterations for three different $\NcellHI$ values.  }
    \label{fig:logA_Afr}
\end{figure}

\begin{figure}
    \centering
    \includegraphics[width= 0.5\textwidth]{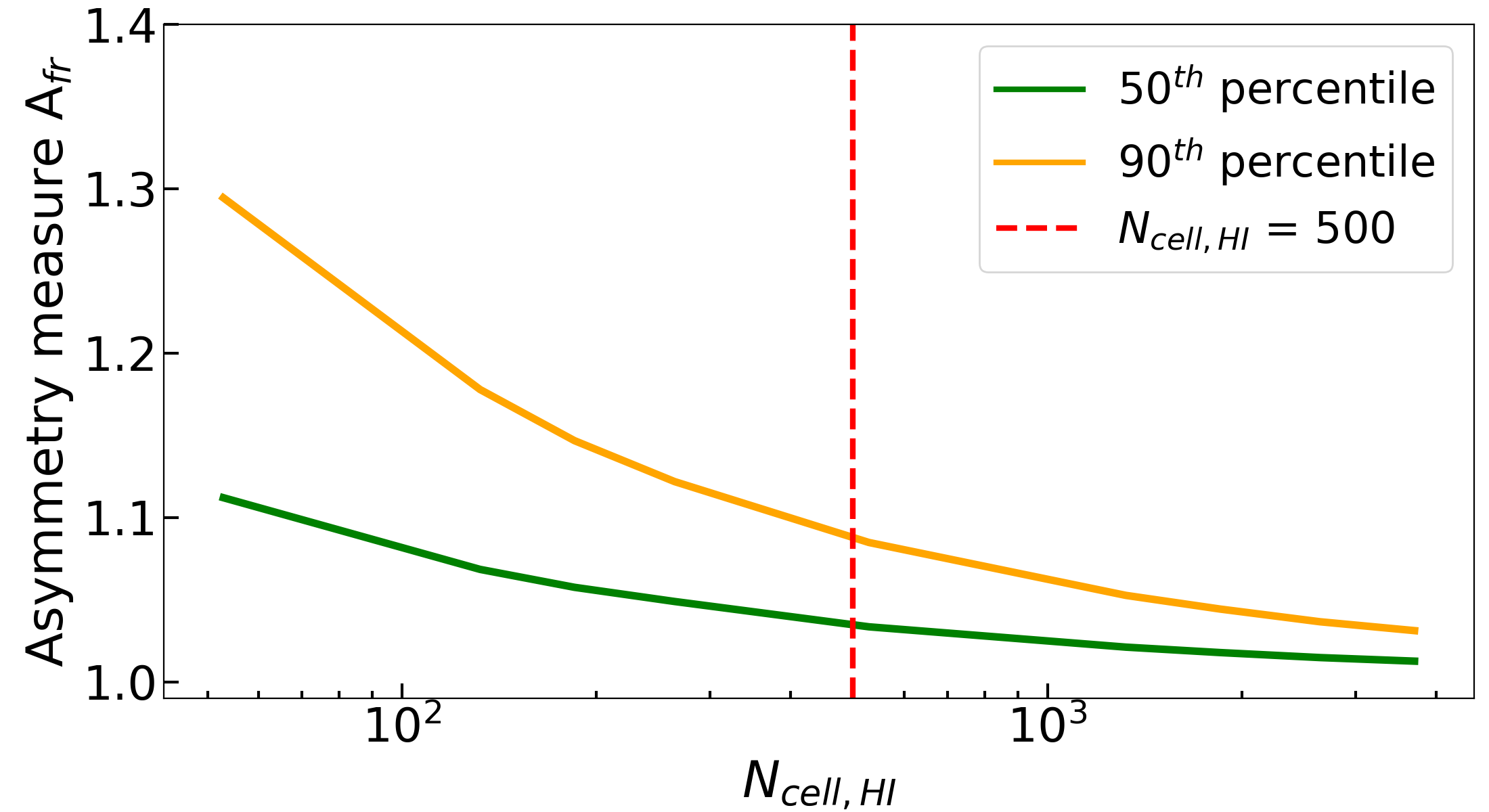}
    \caption{$\Afr$ distribution width as a function of $\NcellHI$.  The 50$^{\rm th}$ (green) and 90$^{\rm th}$ (orange) percentiles of the $\Afr$ distribution as a function of $\NcellHI$ for an intrinsically symmetric TNG100 galaxy. The vertical dashed red line is our adopted quality cut of 500 $\NcellHI$, corresponding to a minimum \HI\ mass of $7\times10^8\ \Msun$.}
    \label{fig:Npart_Afr}
\end{figure}

The fixed mass and consequently spatial resolution of a simulation affects the ability to correctly measure the asymmetry. If the gas distribution is inadequately sampled, the global \HI\ spectrum will not reflect the true kinematics and the asymmetry measurement will be incorrect. To investigate the uncertainty in an $\Afr$ measurement due to the number of contributing gas cells, we adopt a similar method as \ct{watts20} for investigating the effect of measurement noise on $\Afr$. Using a globally \HI-symmetric ($\Afr=1$) TNG100 galaxy consisting of 14198 gas cells with a total \HI\ mass of $\lgMHI = 9.73$, we randomly select 10$^4$ sub-samples of size $\Ncell$ for $\Ncell$ values between 50 and 10$^4$. We compute the spectrum for each sub-sample and measure $\Afr$ and the total \HI\ mass for each spectrum using the same measurement limits as the true line profile. To account for the fact that cells have different \HI\ masses, and that our measurement limits include a subset of all cells in the galaxy, we define the \HI\ mass-weighted cell number as the ratio of the measured \HI\ mass to the nominal cell mass (baryonic mass resolution) of the simulation: $\NcellHI = M_\text{\HI}/1.4\times10^6\, \Msun$.

In the left panel of Fig. \ref{fig:logA_Afr} we show normalised histograms of $\log(A)$ (eq. \ref{eq:A}) for three different $\NcellHI$ values. The distributions are Gaussian with mean $\log(A)$ = 0 and a scatter that increases with lower $\NcellHI$. That is, the effect of inadequate sampling of cells is consistent with random Gaussian noise. This is also reflected in the right panel of Fig. \ref{fig:logA_Afr} where we show the cumulative $\Afr$ distributions of the same samples. The distribution reaches a given cumulative fraction at a higher $\Afr$ value when $\NcellHI$ is lower, and demonstrates that an intrinsically symmetric global \HI\ spectrum will be measured to have $\Afr$ > 1 more frequently than it will $\Afr$ = 1. As this is analogous to what \ct{watts20} found for observational measurement noise, we refer to this effect as `sampling-induced asymmetry'. We must be conscious of this effect when comparing the $\Afr$ distributions of different samples, because if they have different $\NcellHI$ distributions they will have different degrees of sampling-induced asymmetry.

To demonstrate this in a more continuous fashion, in Fig. \ref{fig:Npart_Afr} we show the 50$^{\rm th}$ and 90$^{\rm th}$ percentiles of the $\Afr$ distribution as a function of the mean $\NcellHI$ for each set of $10^4$ sub-samples. To provide a reasonable compromise between the confidence in our $\Afr$ measurements and the number of galaxies studied in this work, we selected galaxies with $\NcellHI \geq 500$ as shown by the red dashed line in Fig. \ref{fig:Npart_Afr}.  
This translates to a minimum \HI\ mass of $M_\text{\HI}=7\times10^{8} M_{\sun}$, {and 5604 galaxies had \HI\ masses below this threshold leaving 10699 galaxies in our final sample.}

\subsection{Final sample properties} \label{subsec:samp}

\begin{figure*}
    \centering
    \includegraphics[width=\textwidth]{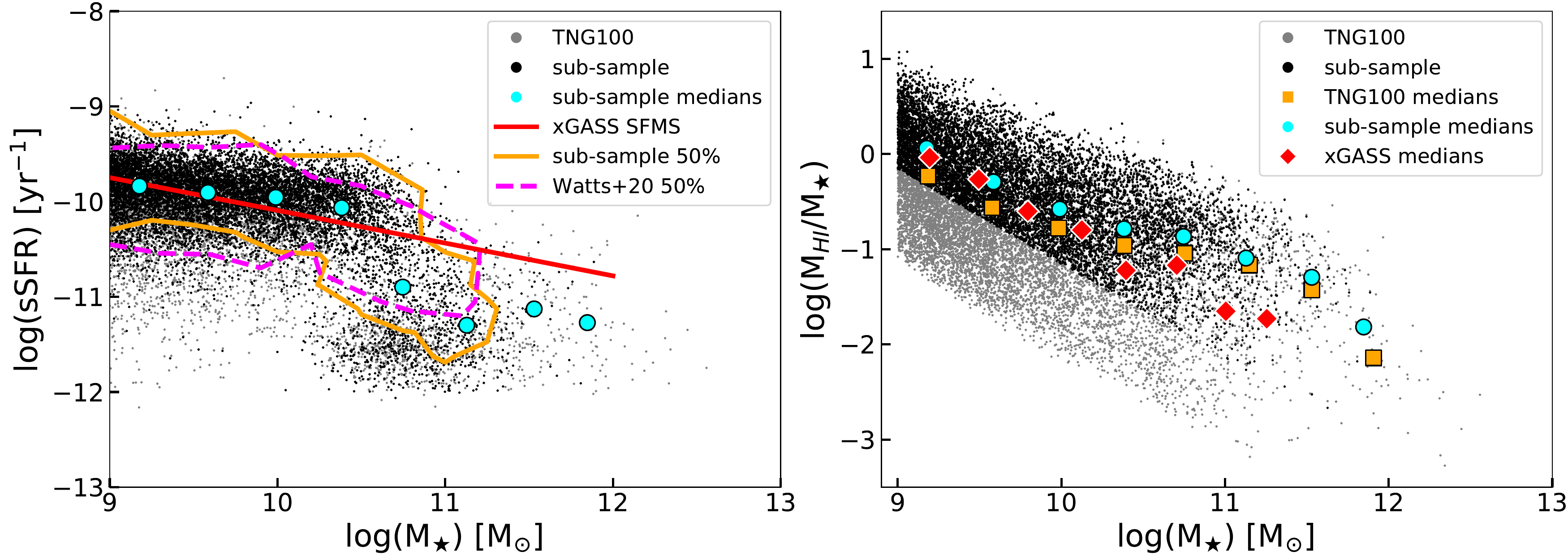}
    \caption{Properties of our final TNG100 sub-sample. In both panels the TNG100 parent sample and our sub-sample are shown as grey and black points, respectively, and median y-axis quantities in bins of 0.4 dex in stellar mass are shown as cyan points. In the left panel we show the ${\rm M}_{\star}-$sSFR plane, with the xGASS SFMS overlaid with solid red line. We defined completeness as the ratio of the number of galaxies in the sub- and parent-sample, and show the 50\% completeness contours for our TNG100 sub-sample, and the xGASS sub-sample from  \ct{watts20},  with orange solid and magenta dashed lines, respectively. In the right panel, we show the \HI-stellar mass fraction scaling relation, with the xGASS weighted median gas fractions, {computed including upper-limits}, and TNG100 parent sample medians shown as red diamonds and orange squares, respectively. {All xGASS measurements have been corrected to the TNG100 cosmology}.}
    \label{fig:sampprop}
\end{figure*}

In Fig. \ref{fig:sampprop} we compare our final TNG100 sub-sample ($\NcellHI\geq500$) to the TNG100 parent sample in the ${\rm M}_{\star}$ $-$ specific star-formation rate (${\rm sSFR}\,[{\rm yr}^{-1}] = ({\rm SFR}/{{\rm M}_\star})$) plane and the \HI--stellar mass fraction scaling relation, including a comparison to the xGASS relations \cp{catinella18} and the xGASS sub-sample from \ct{watts20}. 
TNG100 galaxies with $\log({\rm sSFR}) \, [{\rm yr}^{-1}]<-12$ have been assigned a $\log({\rm sSFR})\,[{\rm yr}^{-1}]$ drawn from Normal distribution with $\mu=-11.5$ and $\sigma=0.2$  to mock {the appearance of a red cloud in the presence of a sSFR detection limit}.
We define the completeness of a population as the ratio of the number of galaxies between the sub- and parent-samples in bins of 0.2\,dex for TNG100 and 0.3\,dex for xGASS, for both $\lgMstar$ and $\log({\rm sSFR})\, [{\rm yr}^{-1}]$. 
In the left panel of Fig. \ref{fig:sampprop} the TNG100 sub-sample medians are roughly consistent with the xGASS star-forming main-sequence (SFMS) until $\lgMstar \sim 10.3$ before dropping into the red cloud. 
In the right panel, {we see that the TNG100 parent sample medians agree with xGASS until $\lgMstar\sim10.7$, and above this mass TNG100 galaxies are more \HI-rich.}
{There are many nuances associated with making this comparison accurately, which we have neglected for the sake of simplicity in this paper, but this observation is consistent with \ct{dave20} (see their fig. 9) who also find that TNG100 over-predicts the cold gas in massive galaxies.}
{For detailed analyses of TNG100 HI properties relative to observations, including xGASS, see \ct{stevens19} and \ct{diemer19}.}
{The $\NcellHI$ cut can be seen $\sim 0.3$ dex below the xGASS weighted-medians, and this results in the TNG100 sub-sample having a higher median \HI-mass fraction compared to the parent sample at all stellar masses.}
Clearly our sub-sample is biased toward relatively gas-rich star-forming galaxies below $\lgMstar \sim 10.3$, and traces the gas-rich, star-forming and quenched populations at higher stellar mass. 
This implies that our asymmetry measurements will be tracing perturbations in a population where environmental processes have not yet removed a significant amount of gas, and we consider this in the interpretation of our results in \S\ref{sec:discussion}.

Compared to the xGASS sub-sample from \ct{watts20} our TNG100 sub-sample is less complete to lower sSFR below $\lgMstar \sim 10.3$ and more complete above this mass.
For these reasons we emphasise that we are not trying to reproduce the xGASS selection function or quantitatively compare our results to those of \ct{watts20}, but instead are looking for qualitative agreement or disagreement. Further, if we are to compare populations of galaxies we must ensure they have similar $\lgMstar$ distributions. A bias toward higher stellar mass galaxies will preferentially sample further below the SFMS and lower \HI-fractions (and vice-versa), resulting in a difference in the contribution of environmental processes to each sample.

Last, it is worth mentioning that some of our mock spectra (e.g. Figs. \ref{fig:spec_examples}, \ref{fig:satellites}, \ref{fig:centrals}, and \ref{fig:RPS}) have noticeable flux in the profile `wings' outside the measurement limits.
{Some of this is due to the visualisations being scaled, while some are consistent with the small wings seen in the global \HI\ spectra of high-quality resolved \HI\ observations \cp[e.g.][]{walter08}.}
{But some spectra show excesses that are not seen in observations.}
{This fraction is $\sim10\%$ in galaxies with $\Afr>1.7$ and $\sim2.5\%$ for those with $\Afr<1.01$, and could be due to unphysical aspects of the feedback mechanisms in the simulation creating high-velocity \HI\ complexes, or possibly the absence of observational measurement noise in our data, which would mask this faint emission.} 
Further, the absence of noise means that we may be sensitive to spectra where one half has a dominant peak, while the other half is primarily faint emission. 
Observationally, in these cases the main peak would be preferentially detected and measured as a single, lower-$\Afr$ Gaussian spectrum  instead of the true, higher-$\Afr$ spectrum, as the fainter emission would be lost under the noise.
{For further discussion and examples of this effect see Appendix \ref{app:noise}.}

\section{Results} \label{sec:results}
\subsection{The rate of asymmetries in TNG100} \label{subsec:TNGrate}

\begin{figure}
    \centering
    \includegraphics[width= 0.5\textwidth]{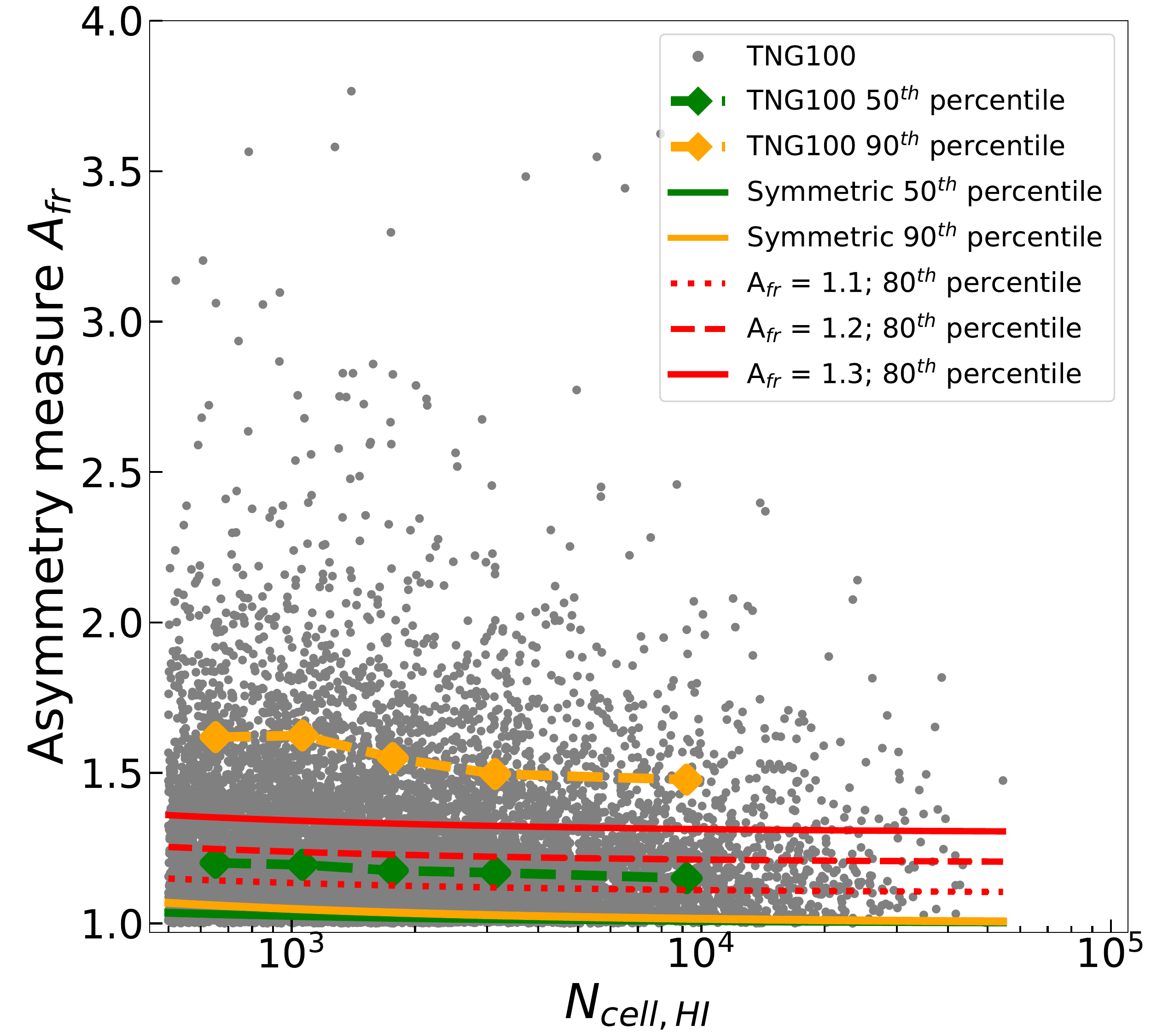}
    \caption{$\NcellHI - \Afr$ parameter space for TNG100. Grey background points correspond to individual TNG100 galaxies, and the dashed green and orange lines are the 50$^{\rm th}$ and 90$^{\rm th}$ percentiles of the distribution of TNG100 galaxies in bins of $\NcellHI$ with centres marked with diamonds. The solid, green and orange lines correspond to the 50$^{\rm th}$ and 90$^{\rm th}$ percentiles for an intrinsically symmetric spectrum from \S\ref{subsec:restest}, respectively. The red dotted, dashed and solid lines are the 80$^{\rm th}$ percentiles for a spectrum with $\Afr=1.1$, 1.2, and 1.3, respectively.}
    \label{fig:Npart_Afr_parameterspace}
\end{figure}

In Fig. \ref{fig:Npart_Afr_parameterspace} we show the $\Afr$ measurements of TNG100 galaxies as a function of their $\NcellHI$, with the 50$^{\rm th}$ and 90$^{\rm th}$ percentiles from \S\ref{subsec:restest} for an intrinsically symmetric spectrum overlaid as solid, green and orange lines respectively. We also overlay the 50$^{\rm th}$ and 90$^{\rm th}$ percentiles of the TNG100 $\Afr$ distribution in five $\NcellHI$ bins designed to have the same number of galaxies in each ({2140}) as dashed lines with diamonds located at the mean $\NcellHI$ in each bin. The TNG100 percentiles are consistently above their corresponding symmetric model percentile, implying that galaxies in TNG100 are more asymmetric than can be attributed to the measurement uncertainty introduced by sampling alone. The result that asymmetries cannot be understood by sampling alone is consistent with results from xGASS \cp{watts20} for observational measurement noise.

To determine the fraction of asymmetric TNG100 galaxies, we must define an $\Afr$ threshold that takes into account the effect of measurement uncertainty introduced by sampling.
{Adopting} the same method as \S\ref{subsec:restest} we modelled the $\NcellHI-\Afr$ parameter-space {using three different projections of a well-resolved ($\NcellHI=10335.1$) galaxy, which we project such that the global \HI\ spectrum has $\Afr=1.1$, $1.2$, and $1.3$.}
{We have also modelled the $\NcellHI-\Afr$ parameter-space using other well-resolved galaxies and can confirm that the model results do not vary between galaxies.}
Asymmetric galaxies are defined as those that have $\Afr$ greater than the 80$^{\rm th}$ percentile of the distribution given their $\NcellHI$ value; i.e. at least 80\% confidence that the $\Afr$ value is not due to sampling-induced asymmetry. 
These thresholds for the three different $\Afr$ values are shown in  Fig. \ref{fig:Npart_Afr_parameterspace} as red dotted, dashed, and solid lines respectively.
The asymmetry rates at each threshold are calculated in the same bins used for the percentiles and for the whole sample and presented in Table \ref{tab:paramspace_rates}.
Using the least asymmetric threshold of $\Afr=1.1$ the rate of asymmetry is relatively constant as a function of $\NcellHI$, in agreement with results  from xGASS \cp{watts20} using the same threshold.
Using the higher $\Afr$ thresholds the rate of asymmetries in the whole sample decreases, but the rate in higher average $\NcellHI$ bins decreases faster with respect to lower average $\NcellHI$ bins.
Assuming that we are correcting for sampling-induced asymmetry correctly, this indicates that galaxies with lower $\NcellHI$, and by construction lower $\MHI$, are more asymmetric.

Interestingly, in using the threshold of $\Afr=1.1$, the {62}\% asymmetry rate in TNG100 is higher than the 37\% observed rate in xGASS. It isn't surprising that the asymmetry rates differ, not only because of the assumptions and approximations that are implicit in the simulated galaxy formation model (e.g. treatment of multiphase interstellar medium), but also because of numerical limitations when discretising the problem (e.g. gravitational softening). We have performed some tests, such as excising the central 1 kpc of galaxies to gauge the impact of softening, and confirm that this does not change our results. Alternatively, we may be more sensitive to spectra with faint emission due to a lack of measurement noise, as mentioned in \S\ref{subsec:samp}, which would reduce the asymmetry rate in observational samples. For these reasons we reiterate here that, when relevant, we present  \textit{qualitative} rather than quantitative comparisons to observational data.

\begin{table}
    \centering
    \caption{Asymmetry rates across the $\NcellHI-\Afr$ parameter space. The average $\NcellHI$ value and the percentage of TNG100 galaxies with $\Afr$ greater than the 80$^{\rm th}$ percentile of three increasing $\Afr$ thresholds in five bins of $\NcellHI$, and for the whole sample (i.e. the average of the 5 bins).}
    \label{tab:paramspace_rates}
    \begin{tabular}{l c c c} \hline
        $\NcellHI$ & $\Afr>1.1$; 80\% & $\Afr>1.2$; 80\% & $\Afr>1.3$; 80\% \\ \hline \hline
        {651.9} & 62\% & 41\% & {29}\% \\
        {1062.3} & {62}\% & {42}\%& 29\%\\
        {1765.1} & {62}\% & {38}\%& {25}\%\\
        {3138.8} & {63}\% & {38}\%& {23}\%\\
        {9208.2} & {60}\%& {36}\%& {21}\%\\ \hline
        Average & {62}\%& {39}\%& {25}\%\\ \hline
    \end{tabular}
\end{table}

\subsection{The relationship between asymmetry and environment} \label{subsec:afr_env}

\subsubsection{Cumulative histogram comparison} \label{subsec:histcomp}
The $\Afr$ distributions of two samples cannot be compared at face value if they have different $\NcellHI$ distributions, as each will have a different level of sampling-induced asymmetry. To make a fair comparison we draw our samples from the same $\NcellHI$ distribution using the same method as \ct{watts20}, which we briefly describe here. We compute the normalised $\NcellHI$ histograms for each sample in bins of 0.1 dex in the range $\NcellHI$ = [500,5000] and treat galaxies with $\NcellHI$ > 5000 as one uniform bin. We define the `common' $\NcellHI$ histogram as the minimum density between the two sample histograms in each bin. The ratio of the common to sample histogram then gives the fraction of galaxies in each bin required to draw the sample from the common distribution. A random uniform variate is drawn for each galaxy and kept if its value is less than this ratio in the relevant bin given by the galaxy's  $\NcellHI$, and we repeat this process $10^4$ times. In addition to this sampling we perform a delete-a-group jack-knife \cp[DAGJK;][]{kott01}  on the dataset to estimate the uncertainty in each bin of the cumulative $\Afr$ histogram. In each of the sampling iterations we  perform five deletions of a random but unique 20\% of the sampled galaxies and compute the cumulative $\Afr$ histogram with each subset removed, and calculate the DAGJK uncertainty estimate for each bin in the cumulative asymmetry histogram. This 20\% is deleted from each bin of the $\NcellHI$ distribution of the sampled galaxies, rather than the whole sample, in order to conserve the shape of the $\NcellHI$ distribution. Finally, we take the average cumulative $\Afr$ value in each bin and the median DAGJK uncertainty in each bin as error bars from the 10$^4$ iterations.

To facilitate a more quantitative comparison between the cumulative $\Afr$ distributions, in Table \ref{tab:asym_rates} we present the fraction of galaxies with $\Afr>1.2$, 1.4, and 1.6, calculated from the cumulative distributions, for each pair of compared samples in the following subsections. These values span the $\Afr$ range where the differences between distributions are typically seen, though we only quote the $\Afr>1.4$ fraction in the text. We stress that these fractions are not `asymmetry rates' corrected for sampling-induced asymmetry as in \S\ref{subsec:TNGrate}, as we are focused on the relative comparison between samples.

\begin{table}
    \centering
    \caption{Relative fractions of the $\Afr$ distributions presented in \S\ref{subsec:afr_env} with $\Afr$ greater than three threshold values. Different figures are separated by solid lines, while samples presented in different panels of the same figure are separated by dashed lines.}
    \label{tab:asym_rates}
    \begin{tabular}{l l c c c} \hline
        Fig. & Sample & & Fraction$>\Afr$ & \\
         &  & $1.2$ & $1.4$ & $1.6$ \\ \hline \hline
        \ref{fig:Afr_sat_cent} & Centrals &{44}\% &{18}\% &7\% \\ 
                                & Satellites &48\% &{22}\% &10\% \\ \hline
        \ref{fig:Afr_halomass}a & $\lgMh \leq 12$ & {45}\% & 18\% & {7}\% \\
                                & $\lgMh = (12,13]$ & {48}\% & {20}\% & {9}\% \\ \hdashline
        \ref{fig:Afr_halomass}b & $\lgMh = (12,13]$ & 49\% & {21}\% & {11}\% \\
                                & $\lgMh = (13,14]$ & {54}\% & {28}\% &{14}\% \\ \hdashline
        \ref{fig:Afr_halomass}c & $\lgMh = (13,14]$ &{52}\% &{26}\% &13\% \\
                                & $\lgMh > 14$ &{50}\% &{27}\% &{15}\% \\  \hline
        \ref{fig:Afr_sats_inout}a &  $R>\Rvir$  &{47}\% & {19}\% & {10}\% \\
                                 &  $R<\Rvir$  & {49}\% &{22}\% &{10}\% \\ \hdashline
        \ref{fig:Afr_sats_inout}b & $R>\Rvir$  &{47}\% &23\% &{12}\% \\
                                 & $R<\Rvir$  &57\% &{31}\% &15\% \\\hline
    \end{tabular}

\end{table}

\subsubsection{Satellites vs centrals}

\begin{figure}
    \centering
    \includegraphics[width= 0.5\textwidth]{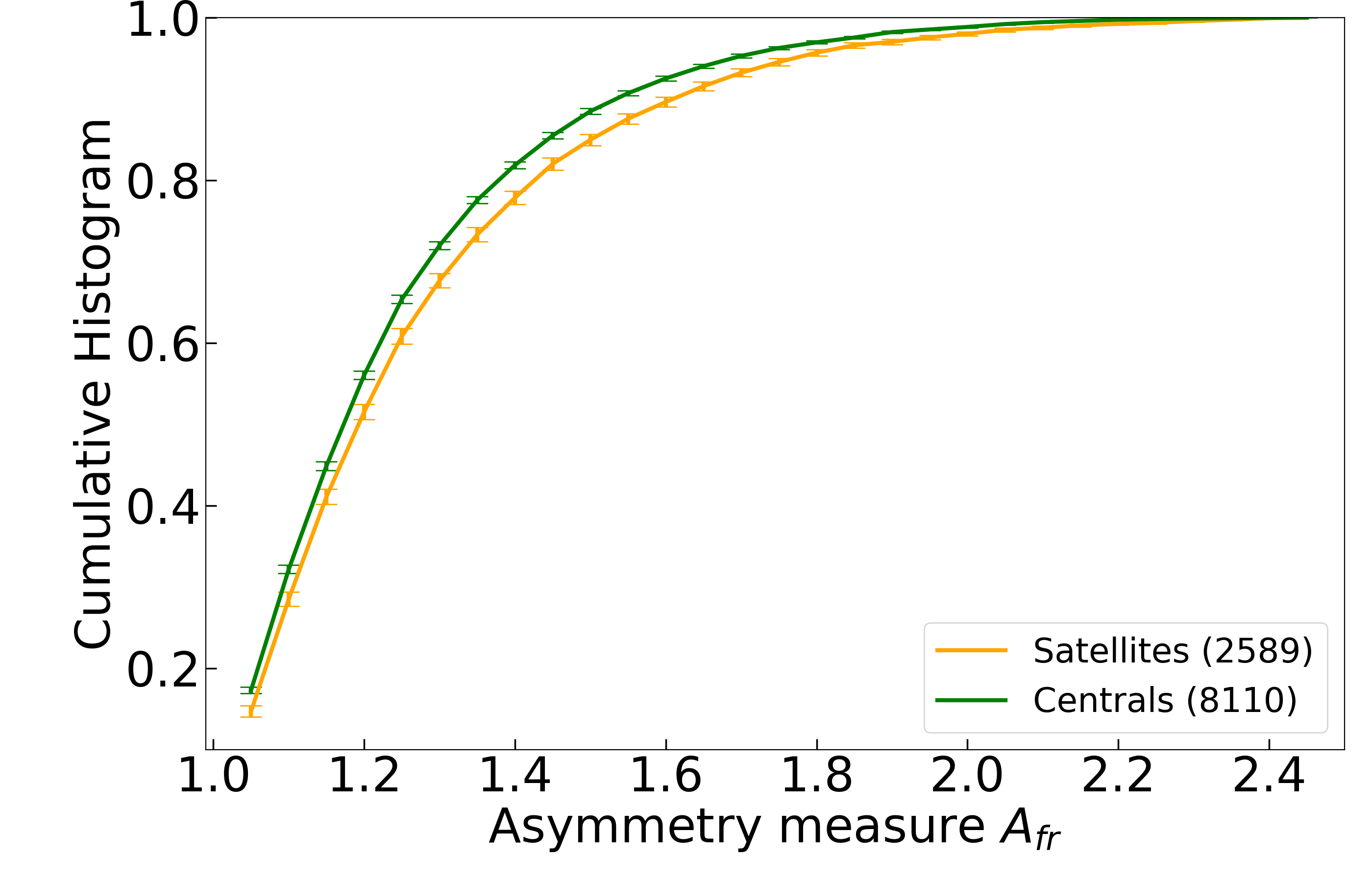}
    \caption{Mean cumulative $\Afr$ distributions of central (green) and satellite (orange) galaxies in TNG100 sampled from the same $\NcellHI$ distribution, with 1$\sigma$ DAGJK error bars. The number of galaxies in each sample is indicated in the legend. }
    \label{fig:Afr_sat_cent}
\end{figure}

If TNG100 represents a reasonable physical model for galaxy formation, then we should see the same environmental trends that we see in the observational data. In Fig. \ref{fig:Afr_sat_cent} we compare the cumulative $\Afr$ distributions of satellite and central galaxies, sampled from the same $\NcellHI$ distribution. The figure reveals that the satellites' cumulative distribution is consistently lower than that of the centrals up to $\Afr=2.2$, at which point they converge, with the largest separation in the $\Afr=[1.3,1.8]$ range. In Table \ref{tab:asym_rates} we see that {22}\% of satellites, compared to {18}\% of centrals have $\Afr>1.4$. This $\sim4\%$ difference is small, but significant, considering the uncertainty in each bin in Fig. \ref{fig:Afr_sat_cent}; hence satellite galaxies, as a population, are more asymmetric than centrals. The exact shape of the cumulative $\Afr$ distributions differ from what \ct{watts20} found in xGASS, but the same qualitative result remains. We have also tested this result by mocking the uncertainties in satellite and central classification typically seen in observational group finding algorithms. To adopt an uncertainty similar to the \ct{yang07} group catalogue used by \ct{watts20}, we randomly reassigned 30\% of centrals as satellites, and vice-versa, and find that satellite galaxies remain a more asymmetric population.

\subsubsection{$\Afr$ distributions as a function of halo mass} \label{subsec:Afrhalo}

\begin{figure}
    \centering
    \includegraphics[width= 0.5\textwidth]{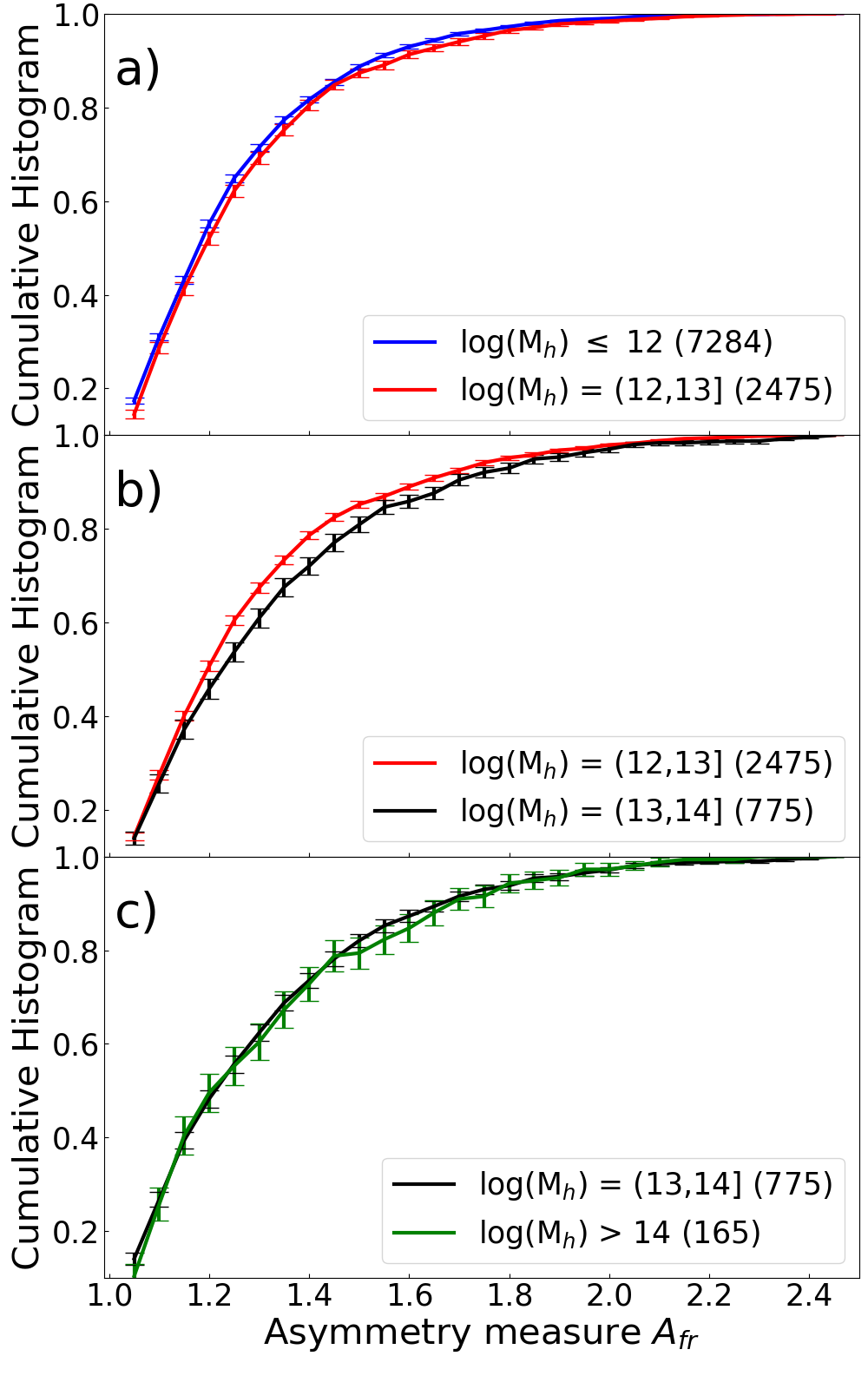}
    \caption{Mean cumulative $\Afr$ histograms with 1$\sigma$ DAGJK error bars compared between adjacent halo mass bins, increasing in halo mass from top to bottom. The halo mass ranges and number of galaxies in each bin are shown in the legend of each panel. }
    \label{fig:Afr_halomass}
\end{figure}

The difference between satellites and centrals is useful for broadly separating galaxies based on the types of environmental processes that they undergo, but satellites and centrals undergo different processes depending whether they exist in an under- or over-dense environment. Galaxies in under-dense environments typically undergo {frequent} gravitational interactions such as mergers where only the central survives, while in over-dense environments galaxy merges are more frequent, satellites can undergo hydrodynamical interactions with the intergalactic medium, and galaxies can experience fast gravitational interactions such as harassment. We use host halo mass $\lgMh$ as a proxy for environment as it is related to the amount of hot gas and density of galaxies, and is a correlated with the gas content of satellites \cp[e.g.][]{brown17} making it likely to be a tracer of the physical processes we might expect to be influencing galaxies. We split our sample into four bins of halo mass and compare the cumulative $\Afr$ distributions of adjacent bins. The stellar mass distributions of the satellites and centrals in our sample are well matched, but this is not true for galaxies in the different halo mass bins. In order to fairly compare two halo mass bins (\S\ref{subsec:samp}) we first sample them from the same stellar-mass distribution using bins of 0.25 dex between $\lgMstar = [9,12.5]$ with one iteration of the method described in \S\ref{subsec:histcomp}. We define the common $\lgMstar$ distribution as the minimum in each bin between the density-normalised $\lgMstar$ distributions of each sample, generate a random uniform variate in [0,1] for each galaxy, and keep those with variates lower than the ratio of the common distribution to their density-normalised $\lgMstar$ distribution in their $\lgMstar$ bin. We then compare the cumulative $\Afr$ distributions of these sub-samples. This potentially introduces another source of uncertainty to our analysis, however we repeated this for multiple, different stellar mass sampling iterations and our results do not change.

We compare the cumulative $\Afr$ distributions of galaxies between adjacent halo mass bins in the three panels (a-c) of Fig. \ref{fig:Afr_halomass}. There is a small difference between the samples in Fig. \ref{fig:Afr_halomass}a, {as 20\% of galaxies in $\lgMh = (12,13]$ haloes have $\Afr>1.4$ compared to 18\% of galaxies in $\lgMh \leq12$ haloes (Table \ref{tab:asym_rates}).}  {This small difference is consistent across the $\Afr$ range, but is} not sufficient in the $\Afr$ range required to explain the difference in Fig. \ref{fig:Afr_sat_cent}. We find no difference between the samples in Fig. \ref{fig:Afr_halomass}c as the two distributions and their error bars overlap for all $\Afr$. The largest difference is in Fig. \ref{fig:Afr_halomass}b where the two distributions diverge outside their error bars in the $\Afr=[1.2,1.9]$ range. In the $\lgMh=(13,14]$ sample {28}\% of galaxies, compared to {21}\% in the $\lgMh=(12,13]$ sample, have $\Afr>1.4$ (Table \ref{tab:asym_rates}). Furthermore, this is {approximately} the transition in halo mass where environmental processes that affect satellites and are explicitly linked to halo mass, such as ram-pressure stripping (RPS), are known to begin to become effective \cp[e.g.][]{catinella13,brown17}. This interpretation is further reinforced by checking the fraction of satellite galaxies in each halo mass bin:  7\%, {49}\%, {92}\% and 100\%, respectively. There is a similar increase in the satellite fraction ($\sim$42\%) between the halo bins in Fig. \ref{fig:Afr_halomass}a  as there is in Fig. \ref{fig:Afr_halomass}b. If a property exclusive to satellite galaxies was the cause of the difference in cumulative $\Afr$ distributions we might expect a similar difference in Fig. \ref{fig:Afr_halomass}a, but we do not see this. Therefore, the cause is more likely to be linked to the transition in halo mass and the change in associated processes. It is worth noting that in TNG100 haloes with $\lgMh>13$, environmental processes make a significant fraction of satellites devoid of gas \cp[][Stevens et al., submitted]{diemer19}. This could explain why we see no difference between the samples in Fig. \ref{fig:Afr_halomass}c, as our data are not sensitive to the satellites that have been strongly affected by environment.

A final observation from Fig. \ref{fig:Afr_halomass} is that even in the lowest halo mass bins, a large fraction ($\sim${45}\%) of the sample has $\Afr>1.2$. This demonstrates, along with Fig. \ref{fig:Afr_sat_cent}, that on average central galaxies are not a global \HI\ symmetric population. This is consistent with observations \cp[e.g.][]{haynes98,espada11,watts20} and we discuss this further in \S\ref{subsec:interpret}.

\subsubsection{$\Afr$ distributions inside and outside $\Rvir$}

\begin{figure}
    \centering
    \includegraphics[width=0.48\textwidth]{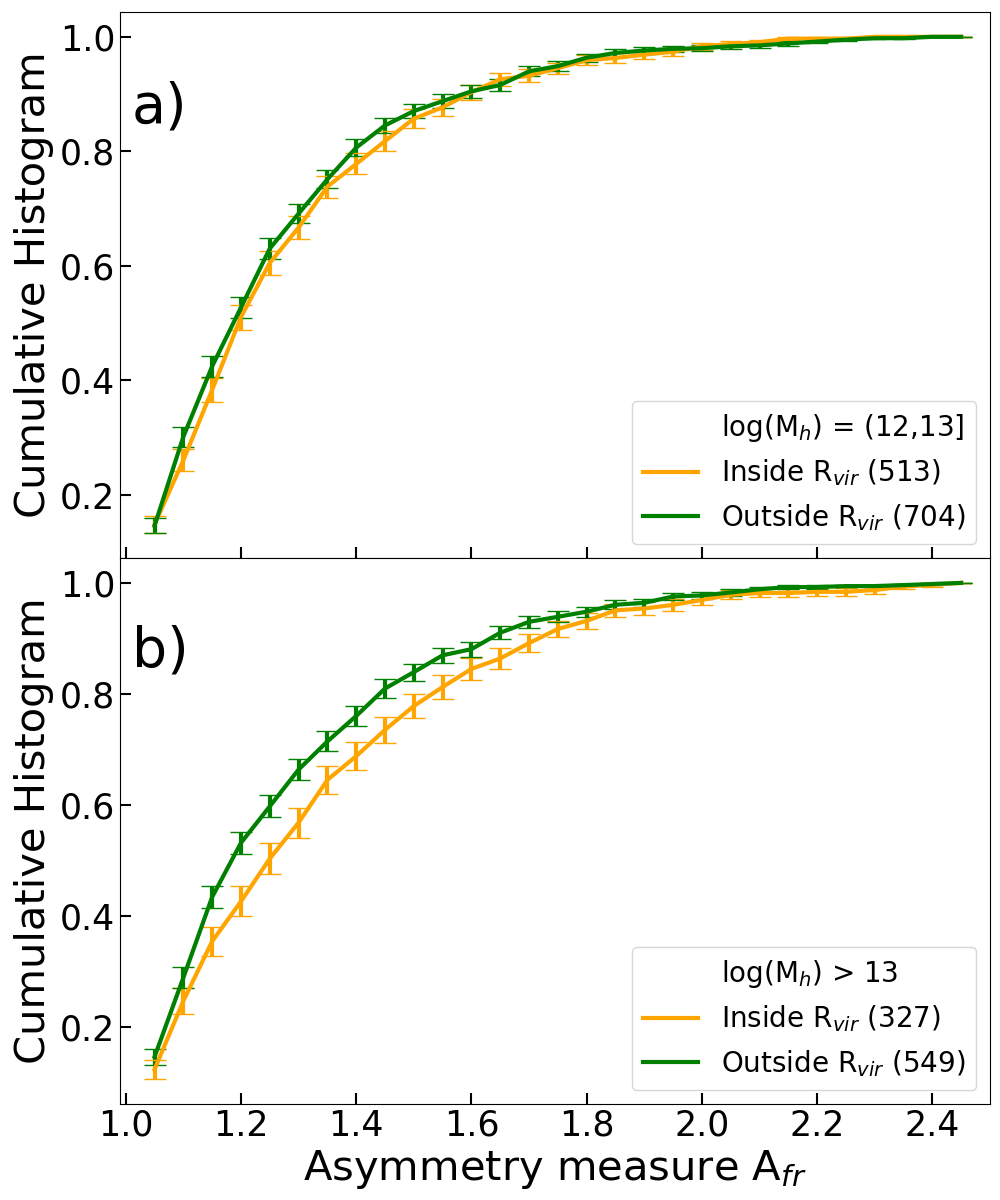}
    \caption{Mean cumulative $\Afr$ distributions of satellite galaxies inside (orange) and outside (green) their host halo $\Rvir$ in two bins of halo mass: $\lgMh = (12,13]$ (top) and $\lgMh > 13$ (bottom).}
    \label{fig:Afr_sats_inout}
\end{figure}

So far we have grouped galaxies into halo mass bins without consideration of their location with respect to the halo centre. 
If an environmental processes linked to halo mass is responsible for the differences seen in Fig. \ref{fig:Afr_sat_cent} and Fig. \ref{fig:Afr_halomass}, we might expect that it would be dominant within the halo's virial radius ($\Rvir$).
In Fig. \ref{fig:Afr_sats_inout} we compare the cumulative $\Afr$ distribution of satellite galaxies inside and outside $\Rvir$ for the $\lgMh =(12,13]$ halo mass bin (panel a), and a combined $\lgMh > 13$ bin (panel b) which we combine to increase statistics as the two samples have no difference in Fig. \ref{fig:Afr_halomass}c. 
The hard radius cut of $\Rvir$ means that the $R>\Rvir$ population will be contaminated to a degree by splash-back galaxies, which can also show disturbed gas morphologies due to ram-pressure stripping \cp{marasco16}. 
However it also contains first in-fall galaxies, {and galaxies identified by {\sc subfind} out to 6 $\Rvir$  within the main friends-of-friends linked halo.}
{These systems provide a control population unaffected by the denser environment inside $\Rvir$.}
We have confirmed that our results do not change when using 0.8$\Rvir$ or 1.2$\Rvir$ instead. 

In Fig. \ref{fig:Afr_sats_inout}a the cumulative distributions and their error bars overlap for all $\Afr$, thus, there is no significant increase in global \HI\ asymmetries for galaxies inside $\Rvir$ in this mass range. 
However, in Fig. \ref{fig:Afr_sats_inout}b the two distributions diverge outside their error bars in the $\Afr=[1.1,1.7]$ range. 
{31}\% of galaxies inside $\Rvir$ have $\Afr>1.4$ compared to 23\% outside $\Rvir$ (Table \ref{tab:asym_rates}), thus galaxies inside $\Rvir$ in $\lgMh > 13$ haloes are, as a population, more asymmetric than those outside $\Rvir$. 
This difference is over a similar $\Afr$ range as observed in Fig. \ref{fig:Afr_halomass}b and the largest separation in Fig. \ref{fig:Afr_sat_cent}, providing further evidence that the perturbing mechanism is environmental in nature, and we investigate this further in \S\ref{subsec:resgas}. 
We have checked the $\lgMstar$ distributions of the galaxies inside and outside $\Rvir$ in both halo mass bins, and find no differences that could be sufficient to explain the difference in cumulative $\Afr$ distributions. 

\subsection{Results summary}
There appears to be a trend where higher asymmetries are linked to denser environment, revealed by tracking the behaviour of the global asymmetry parameter $\Afr$ as a function of different environment metrics. The trends we have found between satellites and centrals, with halo mass, and cluster-centric distance suggest that a hydrodynamical interaction such as RPS could be an important factor. Furthermore, it is clear that global \HI\ asymmetries are often present in galaxies regardless of their environmental classification. We wish to understand if there is a characteristic shape to the global \HI\ spectra of  galaxies that could be related to the cause of their disturbances.

\section{Discussion} \label{sec:discussion}

\subsection{Global \& resolved \HI\ observations and environment} \label{subsec:interpret}

\subsubsection{Global \HI\ spectra of satellites and centrals}
\begin{figure*}
    \centering
    \includegraphics[width= 0.95\textwidth]{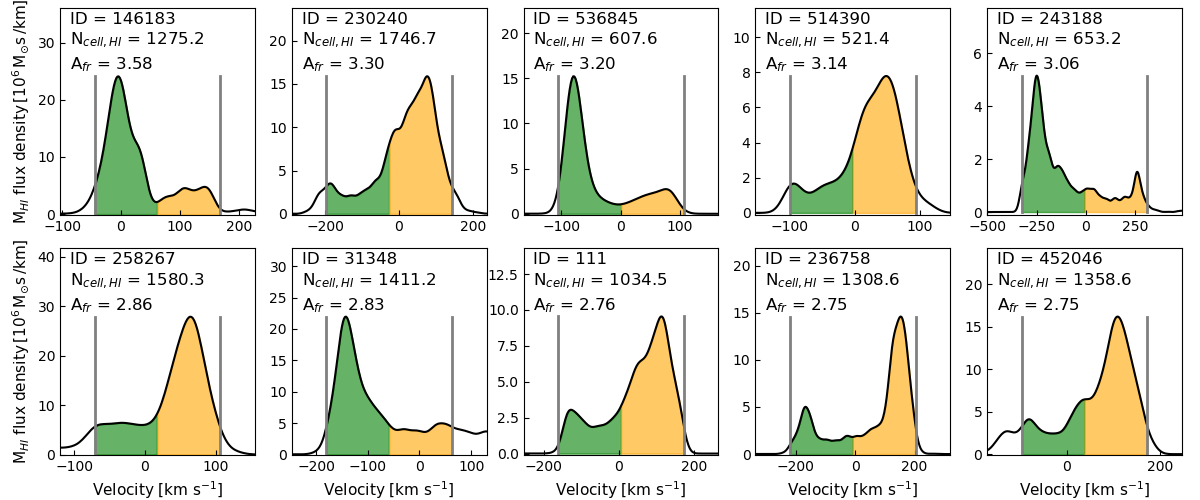}
    \caption{Global \HI\ spectra for the 10 most asymmetric satellite galaxies in TNG100 ordered by decreasing $\Afr$. The legend in each panel indicates the subhalo ID, $\NcellHI$, and $\Afr$ measurement of each galaxy. The velocity axis of each spectrum has been scaled to 1.5$\times \wtw$, and the $\wtw$ measurement limits are shown with grey lines. The integrated flux in each half of the spectrum used to calculate $\Afr$ is shaded green and orange.}
    \label{fig:satellites}
\end{figure*}

\begin{figure*}
    \centering
    \includegraphics[width= 0.95\textwidth]{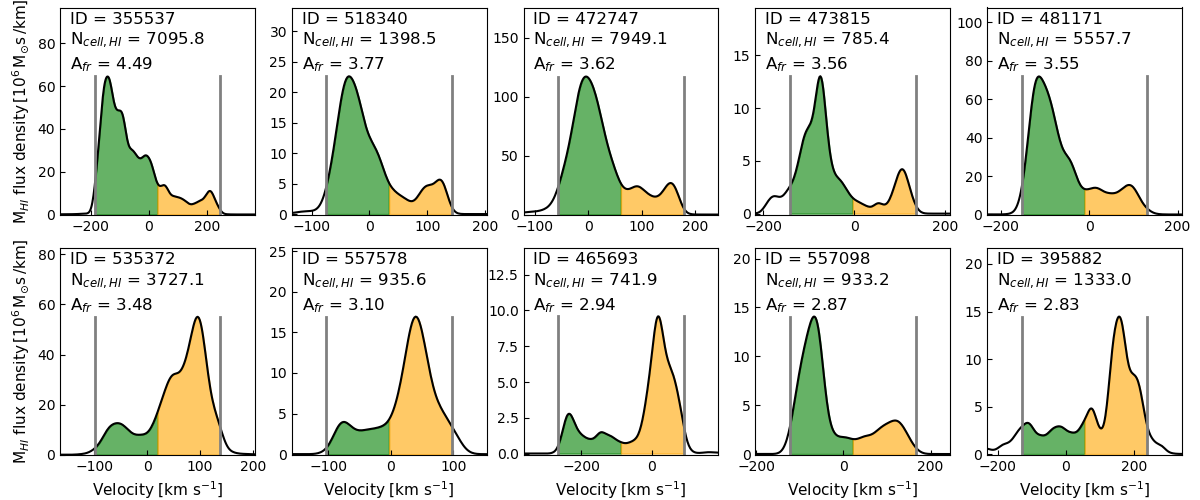}
    \caption{The same as Fig. \ref{fig:satellites} but for the 10 most asymmetric centrals.}
    \label{fig:centrals}
\end{figure*}

We are particularly interested to see if there is a difference in the shape of global \HI\ profiles between centrals, i.e. the most massive subhalo hosted by a dark matter halo; and the less massive satellites that orbit in the halo. 
In Fig. \ref{fig:satellites} and Fig. \ref{fig:centrals} we show global \HI\ spectra for the 10 most asymmetric satellite and central galaxies in our sample, respectively. 
{Interestingly the three highest $\Afr$ galaxies are centrals, which we investigate in \S\ref{subsec:resgas}.}
{Otherwise,} qualitatively there are no distinguishing features between either of the samples: they are both dominated by spectra with very uneven horns or where one horn has been removed. 
Both samples have examples of galaxies with spectra where 0$ \,\kms$ is centred on a peak rather than the centre of the spectrum or near $V_M$ (the overlap of the green and orange shading), {and this is also discussed \S\ref{subsec:resgas})}. 

This is unsurprising and reflects more on the asymmetry measurement that we have adopted and the type of profile shape that it is sensitive to, i.e. where emission is suppressed over half the spectrum rather than small, local variations, instead of any property of satellites or centrals.
We have also confirmed that none of the spectra shown are from face-on systems which also show narrow Gaussian-shaped profiles that can look similar to a single-horn spectrum, but are harder to interpret in terms of a global asymmetry in a rotating disc.

\subsubsection{Resolved gas distributions of satellites and centrals} \label{subsec:resgas}
\begin{figure*}
    \centering
    \includegraphics[width= 0.95\textwidth]{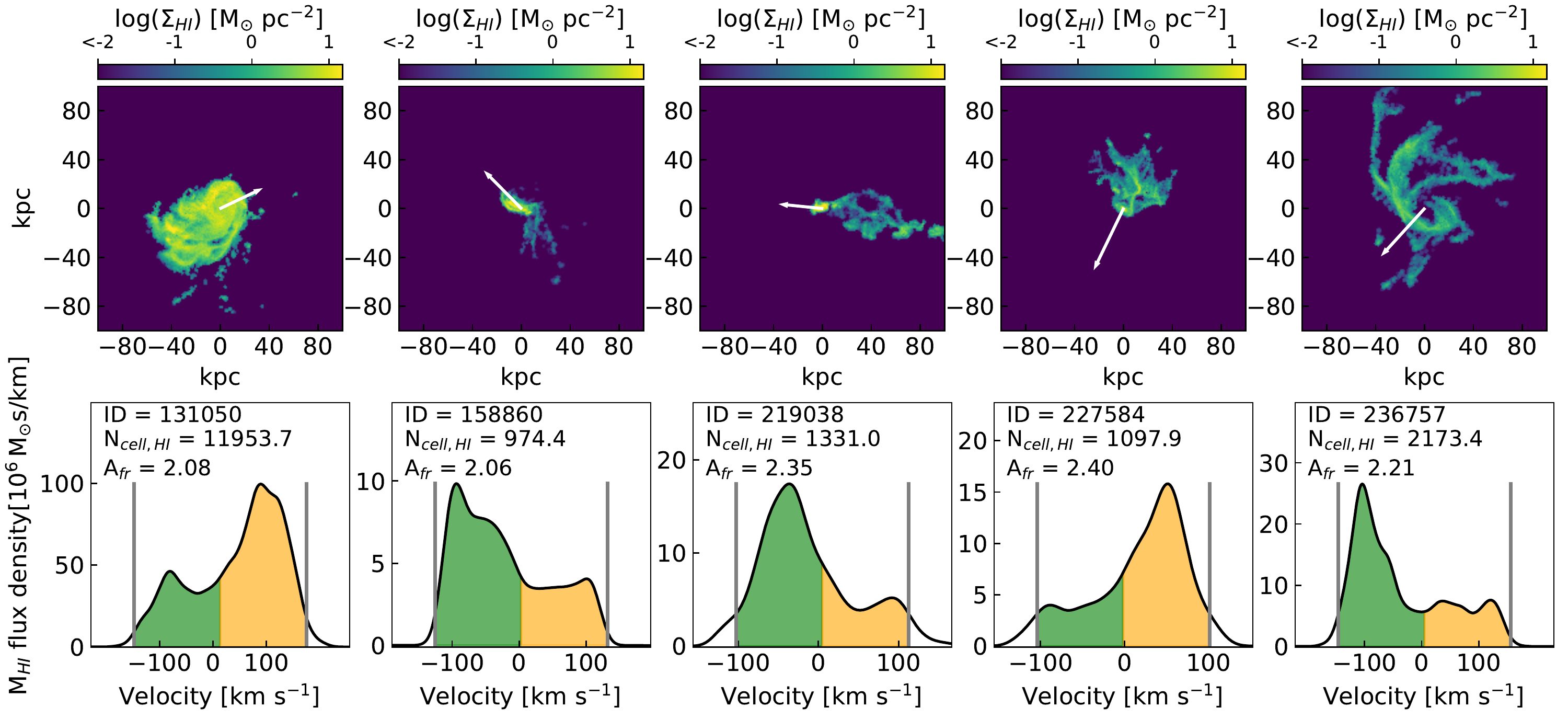}
    \caption{Satellites identified as undergoing RPS. Top row: Resolved \HI\ surface density in the in-box projection, with the density scale given by the colorbar. White arrows indicate the projected velocity vector of each galaxy normalised to a length of 50 kpc. Bottom: Global \HI\ spectrum of each galaxy, shown as in Fig. \ref{fig:satellites}.}
    \label{fig:RPS}
\end{figure*}

The suppression or removal of one horn in a spectrum is typically associated with/seen in satellite galaxies undergoing RPS from the intergalactic medium in galaxy clusters \cp[e.g.][]{kenney04, elagali19}.
However, similar shapes have also been seen in centrals in lower density environments \cp[e.g. NGC 6949;][]{boomsma08}, with the cause attributed to either gas accretion or tidal interactions. 
Does this imply that our interpretation, that a hydrodynamical interaction drives the differences we see, is incorrect (or at least too simplistic)? 
Furthermore, does the gas in central galaxies show the same morphological features as satellites galaxies with the same profile shape, and is there a particular process that is responsible?

We inspected the resolved \HI\ surface density distribution of the 50 most asymmetric satellite and central galaxies in our sample, in attempt to gain insight into the above questions.
{Out of the three centrals {with} greater $\Afr$ than any satellite galaxy in the simulation, two have significant extra-planar \HI\ associated with the position of a satellite galaxy with mass ratio $>0.3$.}
{The one remaining system shows a similar morphology but no satellite galaxy, implying that it could have already been accreted onto the central.}
{This type of gas morphology is also shared by the spectra where 0$ \,\kms$ is centred on a peak, as mentioned above, and observationally these systems would be classified as \HI-confused and likely removed from the sample.}
{Regardless, the highest $\Afr$ measurements in centrals appear to be due to strong tidal interactions or the accretion of \HI\ from an interacting companion.}
{However, this is not a process restricted to only centrals, as some satellites show signatures of mergers (see the end of this subsection), but it is responsible for the strong asymmetry seen in these centrals. }
{This is consistent with \ct{bok19} who found that galaxies in close pairs are more asymmetric than isolated galaxies, but the relationship between asymmetry strength and merging stage remains untested using a sample of ongoing- or post-merger galaxies.}

In the sample of satellites, we conservatively identified galaxies  consistent with what we might assume is representative of the action of RPS, using the criteria that they have a tail of gas that extends in the opposite direction to the galaxy's projected velocity vector. 
We identified five (10\%) satellites with this gas morphology and show them in Fig. \ref{fig:RPS}. 
In contrast, inspection of the sample of centrals indicates that there are no systems that show these morphological features. 
This suggests that a hydrodynamical interaction is responsible for a small fraction of the highest $\Afr$ satellite galaxies in our sample. 
Indeed, this $\sim$10\% is more than sufficient to explain the $4-8$\% (relative) difference in galaxies with $\Afr>1.4$ in the cumulative $\Afr$ distributions seen  in Fig. \ref{fig:Afr_sat_cent}, Fig. \ref{fig:Afr_halomass}b, and Fig. \ref{fig:Afr_sats_inout}b.

\begin{figure*}
    \centering
    \includegraphics[width= 0.95\textwidth]{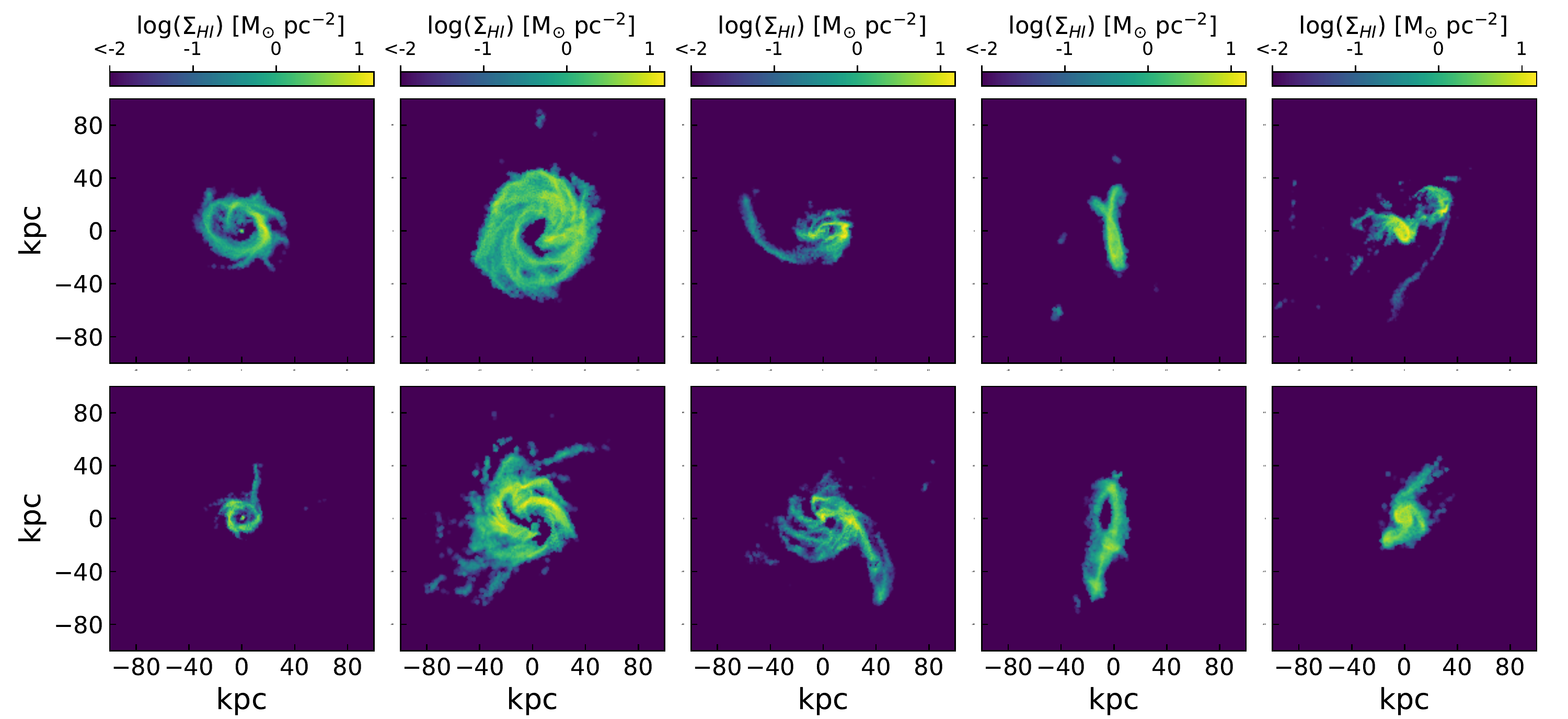}
    \caption{Examples of centrals (top row) and satellites (bottom row) with similar \HI\ morphologies. The resolved \HI\ surface density for the in-box projection is shown, and the density scale is given by the colorbars.}
    \label{fig:sat_cen_examples}
\end{figure*}

This small fraction of galaxies undergoing a hydrodynamical interaction is perhaps unsurprising, as our sample is biased toward gas-rich star-forming systems (\S\ref{subsec:samp}) and not representative of the satellite population, especially in gas-poor environments such as large groups and clusters. A fast-acting process such as RPS \cp[$\sim$ 100 Myr;][]{abadi99}, which first affects the \HI\ that exists primarily outside the stellar disc, would remove a galaxy from our selection quickly and make our sample sensitive to a small fraction of the population truly affected by RPS. Considering this, 10\% of satellites in our sample undergoing a hydrodynamical interaction is consistent with \ct{yun19}, who found that $<$50\% of TNG100 satellites with $\lgMstar>9.5$ and gas fractions $>1$\% in $\lgMh>13$ haloes have a gas morphology consistent with undergoing RPS. This is also consistent with \ct{marasco16}, who found that $\sim 25\%$ of satellites with $\lgMstar>9$ and $\lgMHI > 8.8$  are perturbed by RPS at $z=0$ in the {\sc eagle} Ref-L100N1504 simulation \cp{crain15,schaye15}.

The majority of global \HI\ asymmetries in our TNG100 sample are not driven by a hydrodynamical process, which agrees with observational studies that find that global \HI\ asymmetries, especially strong ones, are present in galaxies regardless of their environment \cp[e.g.][]{espada11,watts20}. If there is a ``ubiquitous" background rate of disturbances in the \HI\ reservoirs of all galaxies, then an environmental process that only affects satellites will build on top of an already perturbed sample, effectively washing out the signature. 
{An alternative explanation could be that gas accretion, which primarily impacts centrals, is responsible {for} the elevation of asymmetry in central galaxies, thus reducing the apparent difference with satellites.}
{The central galaxy population in this work is dominated by objects in $\lgMh<12$ haloes, but studies of gas accretion onto galaxies in these halo masses using {\sc arepo} \cp{nelson13,nelson15} have shown that, at redshift zero, the rate of filamentary accretion is low ($<0.1 \, \Msun\,{\rm yr}^{-1}$) and does not transport material directly to the galactic disc.}
{The simulations of \ct{bournaud05} imply that this accretion rate is insufficient to produce strong global \HI\ asymmetry, but it remains to be quantitatively tested in cosmological simulations, and is worthy of investigation in future work.}

Last, we note that in our sample there is a zoo of gas morphologies that can create the single-horn-like or uneven-peak global \HI\ spectrum, and many of these are shown by both satellites and centrals. In Fig. \ref{fig:sat_cen_examples} we show examples, which include interacting systems such as mergers and streams. There are also galaxies with no clear signs of interaction but substantial holes in the \HI\ distribution, indicating that strong feedback could be sufficient to disturb gas reservoirs in some cases. Hence, the majority of processes that disturb the \HI\ reservoirs of galaxies, at least in our sample, are processes felt by galaxies regardless of being classified as centrals or satellites.

\subsection{Effect of galaxy evolution on the gas reservoirs of galaxies}
In this work we used TNG100 to investigate disturbances in the cold gas reservoirs of galaxies, and find that an environmentally-linked hydrodynamical process can explain the small but higher rate of global \HI\ asymmetries exhibited by satellites compared to centrals. This is consistent with the results of \ct{watts20} who observed a similarly small, but significant, difference between satellites and centrals in a sub-sample of xGASS galaxies, though their sample is biased similarly to ours (\S\ref{subsec:samp}). Regardless of this small difference we find that on average our TNG100 galaxies have \HI\ distributions and global \HI\ profiles that are not-symmetric, regardless of environment, a finding consistent with observations \cp{haynes98,matthews98,scott18,bok19,watts20}. The relative contribution of processes that disturb the gas reservoirs of \textit{all} galaxies thus remains unknown. 

The ubiquity of disturbances in the optical components of galaxies \cp[e.g.][]{zaritsky97,zaritsky13} can be taken as evidence of ubiquitous lopsidedness in the dark matter haloes of galaxies \cp{jog97}. This lopsidedness is also seen in resolved \HI\ observations \cp[e.g.][]{angiras07,vaneymeren11b,reynolds20} and may explain many of the small-amplitude asymmetries in galaxies. Tidal interactions remain a strong candidate \cp[e.g.][]{bok19} for the ubiquity of disturbances in galaxies. Using TNG100, \ct{semczuk20} found that satellite interactions can drive and regenerate the {sigmoidal/}integral-sign shaped warps in \HI\ reservoirs seen in resolved observations \cp{sancisi76}. Further, using controlled simulations, \ct{mapelli08} reproduced the \HI\ reservoir of NGC 891 with a fly-by interaction. Alternatively, the cosmological accretion of gas is expected to be stochastic in the local Universe \cp{sancisi08} and may contribute to disturbances, particularly in the gas-rich regime \cp[e.g.][]{matthews98, bournaud05}. 

It is possible that some of these processes may be identified using additional, or combinations of, asymmetry measures beyond $\Afr$ \cp[e.g.][]{haynes98,matthews98,reynolds20,deg20}. To provide substantial further insight into the effect of galaxy evolution on disturbances in the cold gas reservoirs of galaxies we need the next generation of \HI\ surveys such as The Widefield ASKAP L-band Legacy All-sky Blind Survey \cp[WALLABY;][]{koribalski20}, which is predicted to spatially resolve several thousand galaxies in \HI. However,  blind surveys are biased toward detecting the most gas-rich objects in their volume and resolved data from WALLABY may have limited application in studying the contribution of fast-acting environmental processes, particularly RPS, to disturbances in cold gas reservoirs. Similarly for hydrodynamical simulations, higher mass resolutions than TNG100 are required to make confident global \HI\ asymmetry measurements of galaxies below the median \HI\ scaling relations, and investigate the impact of environment on gas reservoirs in dense environments. This is a regime where the IllustrisTNG 35$/h$ cMpc \cp[TNG50;][]{nelson19,nelson19b,pillepich19} simulation will be able to provide some insight, as it has 2160$^3$ gas cells of nominal mass $8\times10^4\,\Msun$, compared to TNG100's 1820$^3$ cells with $1.4\times10^6\, \Msun$. TNG50 also contains one halo with mass $1.82\times10^{14}\, \Msun$, analogous to the Virgo cluster, giving it the capability to resolve asymmetry in lower \HI\ mass systems in a range of environments. Regardless, cosmological simulations like TNG100 are well suited to testing how spatially resolved asymmetry measurements recover the underlying disturbances in the gas reservoirs of galaxies. This can be used to inform interpretations from applying the same measurements to data from WALLABY, and to test our current models of galaxy evolution in the gas-rich regime.

\section{Conclusions} \label{sec:concl}
In this work we have investigated disturbances in the \HI\ reservoirs galaxies in the IllustrisTNG $75/h$ cMpc box (TNG100) simulation using a sample of {10699} galaxies with $\lgMstarMsun> 9$ and $\lgMHI>8.85$. We have measured the asymmetry of the global \HI\ spectrum as the ratio of the integrated flux in the two halves of the spectrum ($\Afr$) and investigated how this quantity is affected by measurement noise, tracked its behaviour as a function of galaxy environment, and compared it to the resolved gas morphology. Our main results are as follows: 
\begin{itemize}
    \item The typical global \HI\ spectrum in TNG100 is more asymmetric than what can be attributed to measurement uncertainties from stochastic sampling of the \HI\ reservoir. With 80\% confidence we can say that 62\% of the TNG100 galaxies we analysed have $\Afr>1.1$, i.e. an asymmetry greater than 10\%.
    \item  Satellite galaxies, as a population, have more asymmetric global \HI\ spectra relative to centrals, which is qualitatively consistent with observations \cp[see][]{watts20}. Quantitatively, there are 4\% more satellite galaxies with $\Afr>1.4$ relative to centrals.
    \item There is no distinguishing difference between the shape of the global \HI\ spectrum of the most asymmetric satellite and central galaxies, and many satellites and centrals show similar \HI\ morphologies. 
    \item Using halo mass as a proxy for environment we find that the difference between satellites and centrals is likely driven by the satellite population within the virial radius of $\lgMh>13$ haloes, which have {8}\% more galaxies with asymmetry greater than 40\% compared to those outside the virial radius. We speculate that this is due to a hydrodynamical interaction, as this is halo mass where ram-pressure stripping is expected to become effective \cp[e.g.][]{catinella13,brown17,stevens19}. A small fraction of the most asymmetric satellites (10\%) have \HI\ morphologies that resemble a hydrodynamical interaction. This fraction is sufficient to explain the increase in relative asymmetry rates between the populations described above. We conclude that most of the disturbances in the \HI\ reservoirs of galaxies in our sample are primarily driven by processes felt by galaxies regardless of their environment. 
\end{itemize}

It was \ct{richter94} who coined the phrase ``asymmetries in disc galaxies may be the rule, rather than the exception", and it is clear that determining the driver of disturbances in the gas reservoirs of galaxies requires us to go beyond global \HI\ asymmetry measures. However, on a more basic level it is not known whether disturbances in the gas reservoirs of galaxies are primarily present in the distribution or kinematics of the gas, and whether each case is meaningfully traced by global \HI\ asymmetries. Exploring this in the context of environment is not possible with current cosmological simulations, as they are limited to the gas-rich regime when measuring asymmetries in galaxies. However, they remain a useful tool. The creation of mock \HI\ datacubes \cp[e.g.][]{oman19} and measurement of the morphological and kinematic asymmetries adopted in observations \cp[e.g.][]{vaneymeren11a,vaneymeren11b,reynolds20} can lay the groundwork and methodology for determining how global asymmetry measures trace the behavior of baryons in galaxies. Furthermore, the union of this approach with measurements of asymmetry in the stars and ionised gas reservoirs of galaxies \cp[e.g.][]{krajnovic06, bloom18, feng20} might give unique insight into how disturbances propagate \textit{across} and \textit{between} different radii and baryonic phases in galaxies. This is essential to understanding the interplay of the baryonic phases in galaxies and completing our picture of galaxy evolution.


\section*{Acknowledgements}
We extend our thanks to the anonymous referee for their useful comments that improved this paper. ABW acknowledges the support of an Australian Government Research Training Program (RTP) Scholarship throughout the course of this work. LC is the recipient of an Australian Research Council Future Fellowship (FT180100066) funded by the Australian Government. ARHS acknowledges receipt of the Jim Buckee Fellowship at UWA. Parts of this research were supported by the Australian Research Council Centre of Excellence for All Sky Astrophysics in 3 Dimensions (ASTRO 3D), through project number CE170100013.

\section*{Data availability}
The data that support the findings of this study are available upon request from the corresponding author, ABW, and the xGASS and IllustrisTNG data-sets are publicly available at https://xgass.icrar.org/ and https://www.tng-project.org/, respectively.



\bibliographystyle{mnras}
\bibliography{simulationsHIasym_nobold} 



\appendix
\section{Masking faint emission with measurement noise} \label{app:noise}
\begin{figure}
    \centering
    \includegraphics[width= 0.5\textwidth]{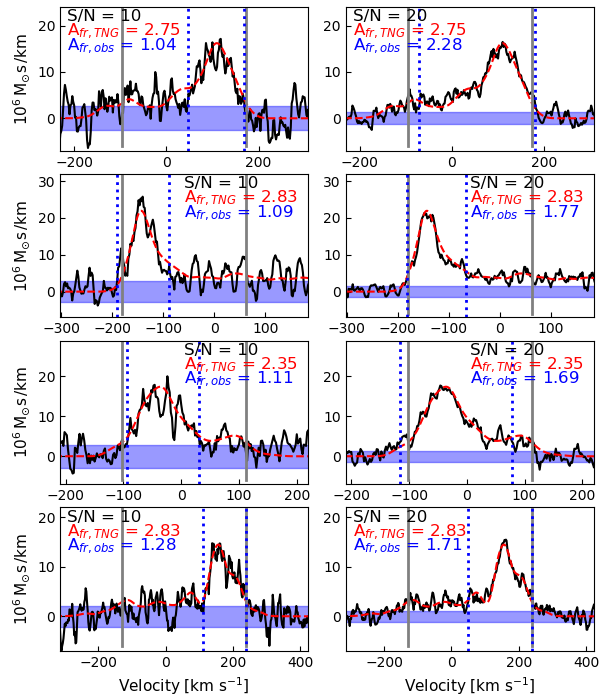}
    \caption{Examples of TNG100 spectra with realistic observational noise added. We show $\SN=10$ (left column) and $\SN=20$ (right column) for four different spectra (each row) with faint, extended \HI\ emission. The intrinsic TNG100 spectrum is overlaid with a red, dashed line, and the TNG100 measurement limits are shown with grey vertical lines. The blue shaded region corresponds to $\pm\rms$, the RMS noise level, and the blue, dotted vertical lines show mock-observational measurement limits. The $\Afr$ value for the TNG100 and mock-observational measurement limits are given at the top of each panel in red and blue, respectively. }
    \label{fig:a1}
\end{figure}

{To investigate the impact of observational measurement noise on the detection of diffuse \HI\ emission, we create mock-observed spectra by adding a random variate drawn from a normal distribution with a mean of zero and $\sigma = \rmsin\, [\Msun\,{\rm s}\,{\rm km}^{-1}]$ to each channel in a spectrum.}
{The magnitude of $\rmsin$ is set by the target integrated signal-to-noise ratio, as parameterised by \ct[][]{saintonge07}}
\begin{equation}
    \SN = \frac{\Sint/\omega}{\rms} \sqrt{\frac{1}{2} \frac{\omega}{\Vsm}},
\end{equation}
{and for simplicity we adopt $\Sint = \MHI$, the spectrum width $\omega$ as defined in \S\ref{subsec:spec}, and we smooth the spectrum with a 5-channel boxcar to a smoothed velocity resolution of $\Vsm=10\, \kms$ to match the typical ALFALFA value.} 
{The required input RMS ($\rmsin$) for a desired $\SN$ spectrum is then given by rearranging eq A1 for $\rms$, the RMS noise in the spectrum after smoothing, and correcting for the boxcar-smoothing noise reduction  $\rmsin = \sqrt{5}\rms$.}
{In Fig. \ref{fig:a1} we show four examples of TNG100 spectra that have extended, faint emission with two representative noise levels corresponding to low/modest ($\SN=10$) and modest/high ($\SN=20$).}
{The noise-inclusive spectrum intersects the $\rms$ level within the TNG100 measurement limits (vertical grey lines) in all the examples regardless of $\SN$, though it happens further from the spectrum's peak at higher $\SN$.} 
{Observationally, this intersection would be sufficient to define the limits of the spectrum, and it is clear that any secondary peaks in the intrinsic spectra are no longer visible.} 
{This demonstrates how this faint emission can be masked by measurement noise.}
{To quantify this, we use these $\rms$-intersection limits to define the mock-observed signal region, and define new measurement limits as where the noise-inclusive spectrum first equals 20\% of it's peak value.}
{Using these new limits, which we show as blue, dotted lines in the Figure, we recompute $\Afr$ and list these values in contrast to the intrinsic TNG100 values at the top of each panel.}
{It is clear that the signal region is always smaller than the intrinsic case, and the decrease in $\Afr$ measurement in all cases reflects the shift toward the measurement of preferentially Gaussian-shaped and more symmetric spectrum. }

\bsp	
\label{lastpage}
\end{document}